\documentclass{aastex}
\usepackage{times}
\usepackage{emulateapj5}

\shortauthors{Bennett et al.}

\shorttitle{MAP Mission}

\newcommand{\iMAP}         {{\it MAP} }
\newcommand{\dg}           {\mbox{$^{\circ}$}}
\newcommand{\lsim}         {\mbox{$_<\atop^{\sim}$}}
\newcommand{\gsim}         {\mbox{$_>\atop^{\sim}$}}

\newcommand{\persec}       {s$^{-1}$}

\slugcomment{The Astrophysical Journal, in press}

\begin{document}

\title{The {\it Microwave Anisotropy Probe} (\iMAP)\altaffilmark{1} Mission}

\author{C. L. Bennett \altaffilmark{2}, 
M. Bay \altaffilmark{3},
M. Halpern \altaffilmark{4},
G. Hinshaw \altaffilmark{2}, 
C. Jackson \altaffilmark{5},
N. Jarosik \altaffilmark{6},
A. Kogut \altaffilmark{2}, 
M. Limon \altaffilmark{2,6}, 
S. S. Meyer \altaffilmark{7},
L. Page \altaffilmark{6},
D. N. Spergel \altaffilmark{8},
G. S. Tucker \altaffilmark{2,9},
D. T. Wilkinson \altaffilmark{6},
E. Wollack \altaffilmark{2},
E. L. Wright \altaffilmark{10}}

\altaffiltext{1}{\iMAP is the result of a partnership between Princeton 
                 University and NASA's Goddard Space Flight Center. Scientific 
		 guidance is provided by the \iMAP Science Team.}
\altaffiltext{2}{Code 685, Goddard Space Flight Center, 
                 Greenbelt, MD 20771}
\altaffiltext{3}{Jackson and Tull, 2705 Bladensburg Road, N.E., Washington, D.C. 20018}
\altaffiltext{4}{Dept. of Physics, Univ. Brit. Col., Vancouver, B.C., Canada V6T 1Z4}
\altaffiltext{5}{Code 556, Goddard Space Flight Center, 
                 Greenbelt, MD 20771}
\altaffiltext{6}{Dept. of Physics, Jadwin Hall, Princeton, NJ 08544}
\altaffiltext{7}{Astronomy and Physics, University of Chicago, 5640 South Ellis Street, 
                 LASP 209, Chicago, IL 60637}
\altaffiltext{8}{Dept of Astrophysical Sciences, Princeton University,
                 Princeton, NJ 08544}
\altaffiltext{9}{Dept. of Physics, Brown University, Providence, RI 02912}
\altaffiltext{10}{Astronomy Dept., UCLA, Los Angeles, CA 90095}

\email{Charles.L.Bennett.1@GSFC.NASA.gov}

\begin{abstract}
The purpose of the \iMAP mission is to determine the geometry, content, 
and evolution of the universe via a 13 arc-min full-width-half-max (FWHM) 
resolution full sky map of the
temperature anisotropy of the cosmic microwave background radiation with
uncorrelated pixel noise, minimal systematic errors, multifrequency 
observations, and accurate
calibration. These attributes were key factors in the success of 
NASA's {\it Cosmic Background Explorer} ({\it COBE}) mission, which 
made a $7^\circ$ FWHM resolution full sky map, discovered temperature anisotropy, and
characterized the fluctuations with two parameters, 
a power spectral index and a primordial amplitude.
Following {\it COBE} considerable progress has been made in higher resolution measurements
of the temperature anisotropy.
With 45 times the sensitivity and 
33 times the angular resolution of the {\it COBE} mission, \iMAP will
vastly extend our knowledge of cosmology. 
\iMAP will measure the
physics of the photon-baryon fluid at recombination.  From this,
\iMAP measurements will constrain models of structure formation, the geometry
of the universe, and inflation. In this paper we present a pre-launch overview of
the design and characteristics of the \iMAP mission.  This information will be 
necessary for a full understanding of the \iMAP data and results, and will also be  
of interest to scientists involved in the design of future cosmic microwave background
experiments and/or space science missions.
\end{abstract}


\section{INTRODUCTION}\label{introcmb}

The existence of the cosmic microwave background (CMB) radiation 
\citep{penzias65}, with its precisely measured blackbody spectrum 
\citep{mather99, mather94, mather90, fixsen96, fixsen94, gush90}, 
offers strong support for the big bang theory.  CMB spatial temperature fluctuations 
were long expected to be present due to large-scale 
gravitational perturbations on the radiation \citep{sachs67}, 
and due to the scattering of the CMB radiation during the recombination 
era \citep{silk68, sunyaev70, Peebles70}.  Detailed computations of model fluctuation power
spectra reveal that specific peaks form as a result of coherent
oscillations of the photon-baryon fluid in the gravitational potential wells
created by total density perturbations, dominated by non-baryonic
dark matter \citep{bond84, bond87, wilson81, sunyaev70, Peebles70}. For a given cosmological
model the CMB anisotropy power spectrum can now be calculated to a high degree 
of precision \citep{hu95, zaldar00}, and since the values of interesting cosmological 
parameters can be extracted from it, there is a strong motivation to measure 
the CMB power spectrum over a wide range of angular scales with accuracy and precision.

The discovery and characterization of CMB spatial temperature 
fluctuations \citep{smoot92, bennett92, wright92, kogut92} confirmed
the general gravitational picture of structure evolution.  The 
{\it COBE} 4-year 
full sky map, with uncorrelated pixel noise, 
precise calibration, and 
demonstrably low systematic errors \citep{bennett96, kogut96a, hinshaw96,
wright96b, gorski96} provides the best constraint on the amplitude
of the largest angular scale fluctuations and has become a standard of 
cosmology (i.e., ``{\it COBE}-normalized'').
Almost all cosmological models currently under active consideration 
assume that initially low amplitude fluctuations in density grew 
gravitationally to form galactic structures.

At the epoch of recombination, $z\sim 1100$, the scattering processes
that leave their imprint on the CMB encode a 
wealth of detail about the global properties of the universe.  
A host of ground-based and balloon-borne experiments have since 
aimed at characterizing these fluctuations at smaller angular 
scales
\citep{
dawson01,           
halverson01,        
hanany00,           
Leitch00,           
wilson00,           
padin01,            
romeo01,            
debernardis00,      
harrison00,         
peterson00,         
baker99,            
coble99,            
dicker99,           
miller99,           
deOliveiraCosta98,  
cheng97,            
hancock97,          
netterfield97,      
piccirillo97,       
tucker97,           
gundersen95,        
debernardis94,      
ganga93,            
myers93,            
tucker93}.          

Multiple groups are presently developing instrumentation and techniques for 
detection of the polarization signature of CMB temperature fluctuations. 
At the time of this writing, only upper bounds on polarization on a variety 
of angular scales have been reported 
\citep{ 
partridge97,           
sironi98,              
torbet99,              
subrah00,              
hedman01,              
keating01}.            

Experimental errors from CMB measurements can be difficult to assess. 
While the nature of the random noise of an experiment is reasonably
straightforward to estimate, systematic measurement errors are not.  None 
of the ground or balloon-based experiments enjoy the extent of systematic
error minimization and characterization that is made possible by a 
space flight mission \citep{kogut92, kogut96a}.  

In addition to the systematic and random errors associated with the
experiments, there is also an unavoidable additional variance associated with
inferring cosmology from a limited sampling of the universe.  A cosmological 
model predicts a
statistical distribution of CMB temperature anisotropy parameters, such as
spherical harmonic amplitudes. In the context of such models, the true CMB
temperature observed in our sky is only a single realization from a
statistical distribution. Thus, in addition to experimental uncertainties,
we account for {\it cosmic variance} uncertainties in our analyses.
For a spherical harmonic temperature expansion $T(\theta ,\phi )=\sum_{\ell
m}a_{\ell m}Y_{\ell m}(\theta ,\phi )$, cosmic variance is approximately
expressed as $\sigma (C_{\ell })/C_{\ell }\approx \sqrt{2/(2\ell +1)}$ where 
$C_{\ell }=<|a_{\ell m}|^{2}>$. Cosmic variance exists independent of the
quality of the experiment. The power spectrum from the 4-year {\it COBE} 
map is cosmic variance limited for $\ell \lesssim 20$.

\begin{figure*}[tb]
\figurenum{1}
\plotone{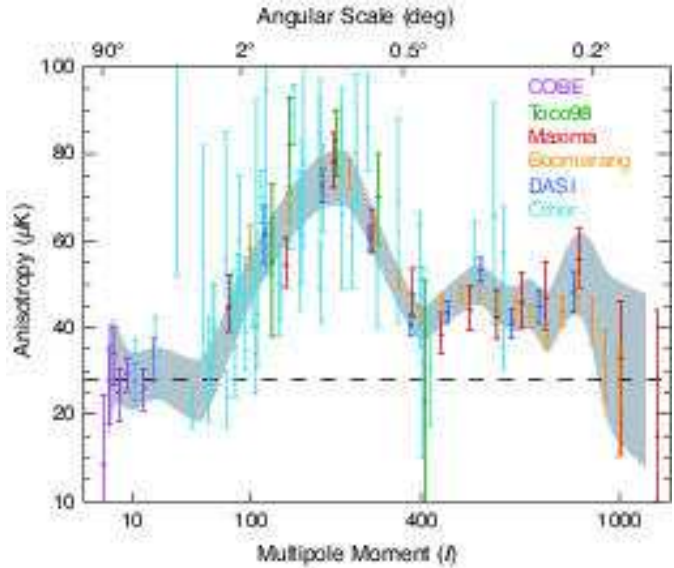}
\caption{The angular power spectrum indicates the state of CMB 
anisotropy measurements at the time of the \iMAP launch.
The grey $2\;\sigma$ band denotes the uncertainty
in a combined CMB power spectrum from recent anisotropy experiments. (Boomerang 
2000 results are omitted in favor of Boomerang 2001 results.)\label{cmbobs}}
\end{figure*}
Fig. \ref{cmbobs} shows the state of CMB anisotropy power spectrum research at about 
the time of the \iMAP launch, 
by combining the results of many recent measurement efforts.  The width of the grey error 
band is determined by forcing the $\chi^2$ of the multi-experiment results to be unity.  
Conflicting 
measurements are thus effectively handled by a widening of the grey band.  Although 
the grey band is consistent with the set of measurements, its absolute correctness is
still entirely dependent on the correctness of the values and errors claimed 
by each experimental group.

\section{COSMOLOGICAL PARADIGMS}\label{introtheory}

The introduction of the inflation model \citep{guth81, sato81, linde82, albrecht83} 
augmented the big bang theory by
providing a natural way to explain why the geometry of universe is nearly flat 
(the ``flatness problem"), why causally separated regions of space share remarkably
similar properties (the ``horizon problem"), and why there is a lack of 
monopoles or other defects observed today (the ``monopole problem'').  While the original inflation model
made strong predictions of a nearly perfectly flat geometry and equal 
gravitational potential fluctuations at all spatial scales, 
inflationary models have since been seen 
to allow for a wide variety of other possibilities.  In its 
simplest conception, the inflaton field that drives inflation is a 
single scalar field.  More generally, a wide variety of formulations 
of the inflaton field are possible, including multiple scalar and non-minimally coupled
scalar fields.  
Thus inflation is a broad class of models, including models that produce an open 
geometry, and models that deviate from generating equal gravitational 
potential fluctuation power on all spatial scales.  
The breadth of possible inflationary scenarios 
has led to the question of whether inflation can be falsified.  CMB observations 
can greatly constrain which inflationary scenarios, if any, describe our universe. 

In the simplest inflationary models, fluctuations arise from adiabatic curvature 
perturbations.  More complicated models can generate isocurvature entropy 
perturbations, or an admixture of adiabatic and isocurvature perturbations.  
In adiabatic models the mass density distribution perturbs the local 
space-time curvature, causing curvature fluctuations 
up through superhorizon scales \citep{bardeen83}.  These are energy density 
fluctuations with a homogeneous entropy per particle.
In isocurvature models the equation of state is perturbed, corresponding 
to local variations in the entropy.  Radiation fluctuations are balanced by 
baryons, cold dark matter, or defects (textures, cosmic strings, global monopole's, 
or domain walls).  Fluctuations of the individual components are 
anticorrelated with the radiation so as to produce no net perturbation in the energy density.
The distinct time-evolution of the gravitational potential between curvature
and isocurvature models lead to very different predictions for the CMB
temperature fluctuation spectrum. These fluctuations carry the
signature of the processes that formed structure in the universe, and of its
large-scale geometry and dynamics. 

In adiabatic models, photons respond to gravitational potential fluctuations due
to total matter density fluctuations to produce
observable CMB anisotropy. The oscillations of the pre-recombination photon-baryon
fluid are understood in terms of basic physics, and their
properties are sensitive to both the overall cosmology and to the nature and
density of the matter.  

The following CMB anisotropy observables should be seen within the context of the 
{\it simplest form} of inflation theory (a single scalar field with adiabatic fluctuations) 
\citep{linde90, kolb90, liddle00}:
(1) an approximately scale-invariant spectral index of primordial fluctuations, 
$n\approx 1$;
(2) a flat $\Omega _{0}=1$ geometry, which places the first acoustic peak 
in the CMB fluctuation spectrum at a spherical harmonic order $\ell \backsim 220$; 
(3) no vector component (inflation damps any initial 
vorticity or vector modes, although vector modes could be 
introduced with late-time defects); 
(4) Gaussian fluctuations with random phases; 
(5) a series of well-defined peaks in the CMB power spectrum, with the 
first and third peaks enhanced relative to the second peak \citep{hu96}; and 
(6) a polarization pattern with a specific orientation with respect to the 
anisotropy gradients.

More complex inflationary models can violate the above properties.  
Also, these properties are not necessarily unique to inflation.
For example, tests of a $n=1$ spectrum of Gaussian fluctuations 
do not clearly distinguish between
inflationary models and alternative models for structure formation.
Indeed, the $n=1$ prediction predates the introduction of the inflation model 
\citep{Harrison70, Zeldovich72, Peebles70}.
Gaussianity may be the weakest of the three tests since the central limit
theorem reflects that Gaussianity is the generic outcome of most statistical 
processes. 

Unlike adiabatic models, defect models do not have multiple acoustic 
peaks \citep{pen97, magueijo96}  
and isocurvature models predict a dominant peak at $l\approx 330$ and a subdominant 
first peak at $l\approx 110$ \citep{hu96}. It is possible to 
construct a model that has isocurvature initial conditions with no superhorizon 
fluctuations that mimics the features of the adiabatic inflationary spectrum 
\citep{turok97}. This model, however, makes very different predictions
for polarization-temperature correlations and for
polarization-polarization correlations \citep{hu97}. By combining temperature anisotropy 
and polarization measurements, there will be a set of tests that are both
unique (only adiabatic inflationary models pass) and sensitive (if the model fails
the test, then the fluctuations are not entirely adiabatic).

If the inflationary primordial fluctuations are adiabatic, 
then the microwave background
temperature and polarization spectrum is completely specified by the power
spectrum of primordial fluctuations, and a few basic cosmological numbers:
the geometry of the universe ($\Omega_0$, $\Lambda_0$), the baryon/photon 
ratio ($\Omega _bh^2$), the
matter/photon ratio ($\Omega _mh^2$), and the optical depth of the universe
since recombination ($\tau $). If these numbers are fixed to match an 
observed temperature spectrum, then the properties of the polarization
fluctuations are nearly completed specified, particularly for $l>30$ 
\citep{kosowsky99}.
If the polarization pattern is not as 
predicted, then the primordial fluctuations can not be entirely adiabatic.

If a polarization-polarization correlation is found on the few degree
scale, then a completely different proof of superhorizon scale 
fluctuations comes into play. Since polarization
fluctuations are produced only through Thompson scattering, then if there
are no superhorizon density fluctuations, there should be no superhorizon
polarization fluctuations \citep{spergel97}.

There are two types of polarization fluctuations: ``E modes'' 
(gradients of a field) and ``B modes'' (curl of a field). Scalar 
fluctuations produce only E modes, while tensor (gravity wave) 
and vector fluctuations produce both E and B modes.
Inflationary models produce gravity waves \citep{kolb90} with a specific
relationship between the amplitude of tensor modes and the slope of the tensor
mode spectrum. The CMB gravity wave polarization signal is extremely weak, 
with a rms amplitude well below 1 $\mu$K. The ability to detect these tracers 
in a future experiment will
depend on the competing foregrounds and the ability to control 
systematic measurement errors to a very fine level. 
Unlike E modes, which \iMAP should detect, the B modes are not correlated 
with temperature fluctuations, so there is no template guide to assist in
detecting these features. Also, the B mode signal is strongest on the largest
angular scales where the systematic errors and foregrounds are the worst. 
The detection of B modes is
beyond the scope of \iMAP; a new initiative will be needed
for a next-generation space CMB polarization mission to address 
these observations.  \iMAP results should be valuable for guiding the design of 
such a mission.

\section{\iMAP OBJECTIVES}\label{object}

The \iMAP mission scientific goal is to answer fundamental cosmology questions
about the geometry and content of the universe, how structures formed, the 
values of the key parameters of cosmology, and the ionization history of the
universe. With large aperture and special purpose telescopes ushering in a new era of
measurements of the large-scale structure of the universe as traced by
galaxies, advances in the use of gravitational lensing for cosmology, the 
use of supernovae as standard candles, and a variety of other  
astronomical observations, the ultimate constraints on cosmological models 
will come from a combination of all these measurements. 
Alternately, inconsistencies that become apparent between observations using 
different techniques may lead to new insights and discoveries.

A map with uncorrelated pixel noise is the most compact and 
complete form of anisotropy data possible
without loss of information. It allows for a full range of statistical
tests to be performed, which is otherwise not practical with the raw data, 
and not possible with further reduced data such as a 
power spectrum.  Based on our experience with the {\it COBE} 
anisotropy data, a map is essential for proper systematic error
analyses.

The statistics of the map constrain cosmological models. \iMAP 
will measure the anisotropy spectral index over a substantial 
wavenumber range, and determine the pattern of peaks.  \iMAP
will test whether the universe is open, closed, or flat
via a precision measurement of $\Omega_{0}$, which is already known to 
be roughly consistent with a flat inflationary universe \citep{knox00}.
\iMAP will determine values of the cosmological constant, the Hubble constant, 
and the baryon-to-photon ratio (the only free parameter in
primordial nucleosynthesis).  \iMAP will also provide an independent 
check of the COBE results, determine whether the anisotropy obeys 
Gaussian statistics, check the random phase hypothesis, and verify whether 
the predicted temperature-polarization correlation is present. \iMAP will 
constrain the inflation model
in several of the ways discussed in \S\ref{introtheory}.
Note that these determinations are independent of traditional astronomical 
approaches (that rely on, e.g., distance ladders or assumptions of virial
equilibrium or standard candles), and are based on samples of vastly 
larger scales.
\begin{table*}[t]
\caption{\iMAP Mission Characteristics \label{tbl-1}}
\small{
\vbox{
\tabskip 1em plus 2em minus .5em
\halign to \hsize {#\hfil &#\hfil \cr
\noalign{\smallskip\hrule\smallskip\hrule\smallskip}
Property & Configuration \cr
\noalign{\smallskip\hrule\smallskip}
Sky coverage              & Full sky \cr
Optical system            & Back-to-Back Gregorian, 1.4 m $\times$ 1.6 m primaries \cr
Radiometric system        & polarization-sensitive pseudo-correlation differential \cr
Detection                 & HEMT amplifiers \cr
Radiometer Modulation     & 2.5 kHz phase switch \cr
Spin Modulation           & $0.464$ rpm $=$ $\sim 7.57$ mHz spacecraft spin \cr
Precession Modulation     & 1 rev hr$^{-1} = \sim 0.3$ mHz spacecraft precession \cr
Calibration               & In-flight: amplitude from dipole modulation, beam from Jupiter \cr
Cooling system            & passively cooled to $\sim 90$ K \cr
Attitude control          & 3-axis controlled, 3 wheels, gyros, star trackers, sun sensors \cr
Propulsion                & blow-down hydrazine with 8 thrusters \cr
RF communication          & 2 GHz transponders, 667 kbps down-link to 70 m DSN \cr 
Power                     & 419 Watts \cr
Mass                      & 840 kg \cr
Launch                    & Delta II 7425-10 on June 30, 2001 at 3:46:46.183 EDT \cr
Orbit                     & $1^\circ - 10^\circ$ Lissajous orbit about second Lagrange point, L$_2$  \cr
Trajectory                & 3 Earth-Moon phasing loops, lunar gravity assist to L$_2$ \cr
Design Lifetime           & 27 months = 3 month trajectory + 2 yrs at L$_2$ \cr
\noalign{\smallskip\hrule}
}}}
\end{table*}
\begin{table*}[t]
\caption{Band-Specific Instrument Characteristics \label{tbl-2}}
\small{
\vbox{
\tabskip 1em plus 2em minus .5em
\halign to \hsize {#\hfil&\hfil#\hfil&\hfil#\hfil
                   &\hfil#\hfil&\hfil#\hfil&\hfil#\hfil\cr
\noalign{\smallskip\hrule\smallskip\hrule\smallskip}
& K-Band\tablenotemark{a} & Ka-Band\tablenotemark{a} &
Q-Band\tablenotemark{a} &
V-Band\tablenotemark{a} & W-Band\tablenotemark{a} \cr
\noalign{\smallskip\hrule\smallskip}
Wavelength (mm)\tablenotemark{b}        & 13  & 9.1 & 7.3 &  4.9 &  3.2 \cr
Frequency (GHz)\tablenotemark{b}        & 23  &  33 & 41  &  61  &  94  \cr
Bandwidth (GHz)\tablenotemark{b,c}      & 5.5 & 7.0 & 8.3 & 14.0 & 20.5 \cr
Number of Differencing Assemblies       &  1  &  1  &  2  &   2  &  4 \cr
Number of Radiometers                   &  2  &  2  &  4  &   4  &  8 \cr
Number of Channels                      &  4  &  4  &  8  &   8  &  16 \cr
Beam size (deg) \tablenotemark{b,d}     &  0.88 & 0.66 & 0.51 & 0.35 & 0.22 \cr
System temperature, $T_{sys}$ (K)\tablenotemark{b,e} 
                                        & 29  &  39 &  59 &  92  &  145 \cr
Sensitivity (mK sec$^{1/2})$  \tablenotemark{b} 
                                        & 0.8  & 0.8 & 1.0 & 1.2 & 1.6 \cr
\noalign{\smallskip\hrule
\tablenotetext{a}{Commercial waveguide band designations used for the
five \iMAP frequency bands.}
\tablenotetext{b}{Typical values for a radiometer are given.   
See text, \citet{page02}, and \citet{jarosik02} for exact values, which
vary by radiometer.}
\tablenotetext{c}{Effective signal bandwidth.}
\tablenotetext{d}{The beam patterns are not Gaussian, and thus are not
simply specified. 
The size given here is the square-root of the beam solid angle.}
\tablenotetext{e}{Effective system temperature of the {\it entire
system}.}}
}}}
\end{table*}
The high-level features of the \iMAP mission are described below and summarized in 
Tables \ref{tbl-1} and \ref{tbl-2}.  
The mission is
designed to produce a full ($>95$\%) sky map of the cosmic microwave background
temperature fluctuations with:
\begin{list}
   {$\bullet$}
   {\setlength{\itemsep}{0 ex}
    \setlength{\topsep}{0.2 ex}
    \setlength{\parsep}{0 ex}}
\item $\approx 0.2^\circ$ angular resolution
\item accuracy on all angular scales $>0.2\dg$
\item minimally correlated pixel noise
\item polarization sensitivity
\item accurate calibration ($<0.5$\% uncertainty)
\item an overall sensitivity level of $\Delta T_{\rm rms}<20~\mu $K per pixel 
        (for 393,216 sky pixels, $3.2\times 10^{-5}$ sr per pixel)
\item systematic errors limited to $<5$\% of the random variance on all
	angular scales
\end{list}

Systematic errors in the final sky maps can originate from a variety of
sources: calibration errors, external emission sources, internal emission
sources, multiplicative electronics sources, additive electronics sources,
striping, map-making errors, and beam-mapping errors.  {\it The need 
to minimize
the level of systematic errors (even at the expense of sensitivity,
simplicity, cost, etc.) has been the major driver of the \iMAP design}.  To
minimize systematic errors \iMAP has:
\begin{list}
   {$\bullet$}
   {\setlength{\itemsep}{0 ex}
    \setlength{\topsep}{0.2 ex}
    \setlength{\parsep}{0 ex}}
\item a symmetric differential design
\item rapid large-sky-area scans
\item 4 switching/modulation periods
\item a highly interconnected and redundant set of differential observations
\item an L$_{2}$ orbit to minimize contamination from Sun, Earth, and 
      Moon emission and allow for thermal stability
\item multiple independent channels
\item 5 frequency bands to enable a separation of galactic and cosmic signals
\item passive thermal control with a constant Sun angle for thermal and power stability
\item control of beam sidelobe levels to keep the Sun, Earth, and Moon levels $<1~\mu$K
\item a main beam pattern measured accurately in-flight (e.g., using Jupiter)
\item calibration determined in-flight (from the CMB dipole and its modulation from MAP's motion)
\item low cross-polarization levels (below -20 dB)
\item precision temperature sensing at selected instrument locations
\end{list}
\begin{figure*}[tb]
\figurenum{2}
\plotone{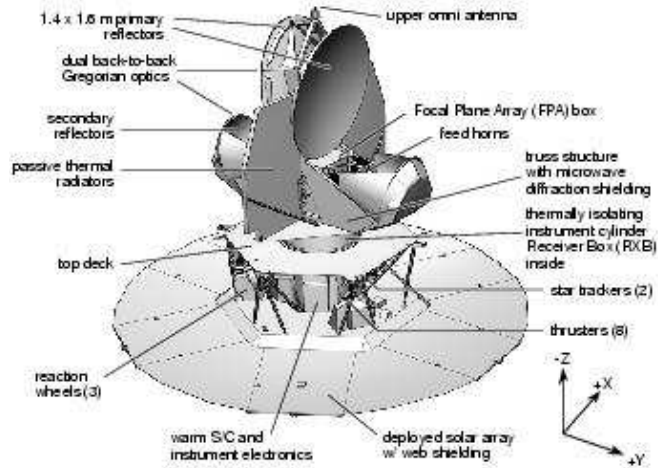}
\caption{A view of the \iMAP Observatory is shown with 
several of the major constituents called out. The Observatory is 3.8 m tall 
and the deployed solar array is 5.0 m in diameter. The Observatory mass is 836 kg.\label{observatory}}
\end{figure*}
\begin{figure*}[tb]
\figurenum{3}
\plotone{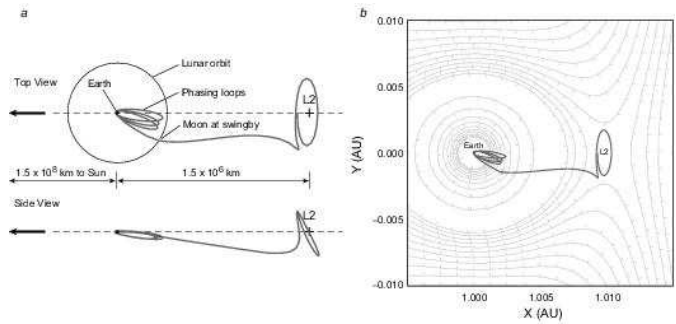}
\caption{Views of \iMAP's trajectory to an orbit about L$_2$.  
\iMAP uses an L$_2$ orbit to enable passive cooling 
and to minimize systematic measurement errors.  
(a) Perspective views (from the North ecliptic pole 
and from within the ecliptic plane) of a typcal trajectory are shown in an Earth 
co-rotating coordinate system.  
An on-board reaction control (propulsion) system 
executes three highly elliptical ``phasing loop" orbits about the Earth, 
which set up a gravity-assist lunar swing-by, and then a cruise to an orbit about 
the second Earth-Sun Lagrange point, L$_2$.  
(b) The co-rotating gravitational potential is shown.  The break in contour lines represent a
change of scale, where the gravitational potential near the Earth is much steeper than near 
L$_2$.  Tick marks indicate the ``down hill'' side of each contour.
The L$_2$ orbit provides a quasi-stable 
orbit in a saddle-shaped gravitational potential.  This is a ``Lissajous'' rather than
``halo'' orbit since the Observatory is at a different position with a different velocity
vector after each six-month loop.\label{trajectory}}
\end{figure*}
\section{HIGH-LEVEL SCIENCE MISSION DESIGN FEATURES}\label{features}
Fig. \ref{observatory} shows an overview of the \iMAP Observatory.  A deployable 
sun shield, with web blankets between solar panels, keeps the spacecraft and 
instrument in shadow for all nominal science operations.  Large passive radiators 
are connected, via heat straps, directly to the High Electron Mobility Transistor 
(HEMT) amplifiers at the core of the radiometers.  A (94 cm diameter $\times$ 33 cm length 
$\times$ 0.318 cm thick) gamma-alumina cylindrical shell provides
exceptionally low thermal conductance (0.59 and 1.4 W m$^{-1}$ K$^{-1}$ 
at 80 K and 300 K, respectively) between the warm spacecraft and 
the cold instrument components.  The back-to-back optical system can be seen as satisfying
part of the requirement for a symmetric differential design.
An L$_2$ orbit was required to minimize thermal variations while simplifying the 
passive cooling design, and to isolate the instrument from microwave emission from
the Earth, Sun, and Moon.  Fig. \ref{trajectory} shows the \iMAP trajectory, including its
orbit about L$_2$.

Further systematic error suppression features of the \iMAP mission are discussed 
in the following subsections. 

\subsection{Thermal and Power Stability}\label{stability}

There are three major objectives of the thermal design of \iMAP.  The first
is to keep all elements of the Observatory within nondestructive temperature
ranges for all phases of the mission.  The second objective is to  
passively cool the instrument front-end microwave amplifiers and reduce the 
microwave emissivity of the front-end components to improve sensitivity. The 
third objective is to use only passive thermal control throughout the entire
Observatory to minimize all 
thermal variations during the nominal 
observing mode.  While the first objective is common to all space
missions, the second objective is rare, and the third objective 
is entirely new and provides significant constraints to the overall thermal 
design of the mission.

All thermal inputs to \iMAP are from the Sun, either from direct thermal 
heating or indirectly from the electrical dissipation of the solar energy 
that is converted in the solar arrays.  Since both thermal changes and
electrical changes are potential sources of undesired systematic  
errors, measures are taken to minimize both.  The slow annual 
change in the effective solar constant is easily accounted for.  
Variations that occur synchronously with the spin period pose the 
greatest threat since they most closely mimic a true sky signal.

To minimize thermal and electrical variations the solar arrays maintain 
a constant angle relative to the Sun of $22.5\dg \pm 0.25\dg$ during CMB
anisotropy observations at L$_2$.  The constant solar angle, combined with 
the battery, provides for a stable power input to all electrical systems.  
Key systems receive further voltage referencing and regulation.

To further minimize thermal and electrical systematic effects, efforts are 
made to minimize variations in power dissipation.  {\it All thermal 
control is passive; there are no proportional heaters and no heaters that switch 
on and off} (except for survival heaters that are only needed in cases of 
spacecraft emergencies, and the transmitter make-up heater, discussed below).  
The electrical power dissipation changes of the various electronics boxes are negligible.  

Passive thermal control required careful adjustment and testing of the thermal blankets, 
radiant cooling surfaces, and ohmic heaters.  A detailed thermal model was used 
to guide the development of a baseline design.  Final adjustments were based on 
tests in a large thermal vacuum chamber, in which the spacecraft (without its solar panels) 
was surrounded by 
nitrogen-cooled walls and the instrument was cooled by liquid helium walls.

Radio frequency interference (RFI) from the transmitter poses a potential 
systematic error threat to the experiment.  Thus, there is a motivation to
turn the transmitter on for only the least amount of time needed to down-link 
the daily data.  However, the power dissipation difference between the 
transmitter on and off states poses a threat to thermal stability.  To  
mitigate these thermal changes, a $53~\Omega$ make-up heater is placed on the transponder 
mounting plate to approximately match the 21 Watt thermal power dissipation difference between 
the on versus off states of the transmitter.  There are two transponders (one is 
redundant) and they are both mounted to the same thermal control plate.  Both 
receivers are on at all times.  The heater can be left on, except for the 
$\approx 40$ minute per day period that the transmitter must be used.  Thus, there 
are two in-flight options for minimizing systematic measurement errors due to 
the transmitter.  Should in-flight RFI from the transmitter be judged 
a greater threat than residual thermal variations, then the transmission time
can be minimized and the make-up heater used.  Alternately, should the residual
thermal variations be the greater threat the transmitter can be left on 
continuously.  The mission is designed such that either 
option is expected to meet systematic error requirements.

There are scores of precision platinum resistance thermometers (PRTs)  
at various locations to
provide a quantitative demonstration of thermal stability at the sub-millikelvin
level.  The information from these sensors is invaluable for making 
a quantitative assessment of the level of systematic errors from residual 
thermal variations and could be used 
to make error corrections in the ground data reduction pipeline, if needed.  The
design is to make these corrections unnecessary and to use the sensor data 
only to prove that thermal variations are not significant.

\subsection{Sky Scan Pattern}\label{skyscan}

The sky scan strategy is critical to achieve minimal systematic 
effects in CMB anisotropy experiments.  The ideal scan strategy would be to
instantaneously scan the entire sky, and then rapidly repeat the scans so 
that sky regions are traversed from all different angles.  
Practical constraints, 
of course, limit the scan rate, the available 
instantaneous sky region, and the angles through which each patch of sky is
traversed by a beam.  For a space mission, increasing the scan speed rapidly
becomes expensive: it becomes more difficult to reconstruct an accurate pointing
solution; the torque requirements of on-board control components increase; and 
the data rates required to prevent beam smearing increase.
The region of sky available for scanning is limited by the acceptable 
level of microwave pick-up from the Sun, Earth, and Moon.
\begin{figure*}[tb]
\figurenum{4}
\plotone{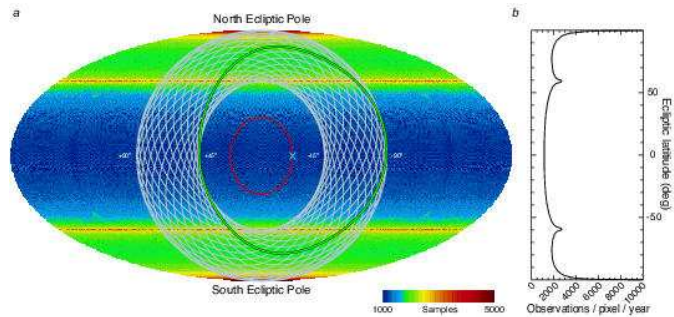}
\caption{(a) A full sky map projection in ecliptic coordinates 
shows the number of independent data samples taken per year by sky position.
Full sky coverage results from the combined motions of the spin, precession, and 
orbit about the Sun.  The spin axis precesses along the red circle in 1 hour.  
When the spin axis is at the position of the blue ``x'' a feed pair traces
the green circle on the sky in the 129 second spin period.  
The white circles indicate the results of the precession.  Full sky coverage is
achieved in six months as \iMAP orbits around the Sun (with the Earth).  
(b) The number of observations per pixel as a function of ecliptic latitude is shown for 
each full year of observations.  The number of observations per pixel will vary by 
frequency band due to differing sampling rates, differing beam solid angles, and data
flagging.  The plot is illustrative; the \iMAP data must be used for exact sky 
sampling values.
\label{scanorbit}}
\end{figure*}

It is possible to quantitatively assess the quality of a sky scanning 
strategy by computer simulation.  Systematic errors are generated as part 
of an input to a computer simulation that converts time-ordered data to
sky maps.  The suppression factor of systematic error 
levels going from the raw time-ordered data into the sky map is a measure of 
the quality of the sky scanning strategy.  A poor scanning strategy will 
result in a poor suppression 
factor.  Our computer simulations show that the sky scanning strategy 
used by the {\it COBE} mission was 
very nearly ideal as it maximally suppressed systematic errors 
in the time-ordered data
from entering the map.  A large fraction of the full sky was scanned 
rapidly, consistent with avoiding a $60\dg$ full-angle cone in the 
solar direction.  However, the Moon was often in and near the beam.  This was 
useful for checks of amplitude and pointing calibration, but much data 
had to be discarded when contamination by lunar emission was significant.  
{\it COBE}, in its low Earth orbit,  also suffered from pick-up of microwave emission from the 
Earth, also causing data to be discarded.

In its nominal L$_2$ orbit the \iMAP Observatory executes a compound spin (0.464 rpm)
and precession (1 hr$^{-1}$), as shown in Figure \ref{scanorbit}.  
The \iMAP sky scan strategy is a compromise.  While 
the \iMAP scan pattern is almost as good as {\it COBE}'s with regard to 
an error suppression factor, it is far 
better than {\it COBE}'s for rejecting microwave signals from
the Sun, Earth, and Moon microwave signals.  

The \iMAP sky scan pattern results in full sky coverage with some variation 
in the number of observations per pixel, as shown in Figure \ref{scanorbit}.

To make a sky map from differential observations, it is also essential for
the pixel-pair differential temperatures to be well interconnected between
as many pixel-pairs as possible.  The degree and rate of convergence of
the sky map solution depends upon it.  The \iMAP sky scan pattern is seen, 
by computer simulation, to enable the creation of maps that converge in a rapid and
well-behaved manner.

\subsection{Multi-Frequency Measurements}\label{multifreq}

\begin{figure*}
\figurenum{5}
\plotone{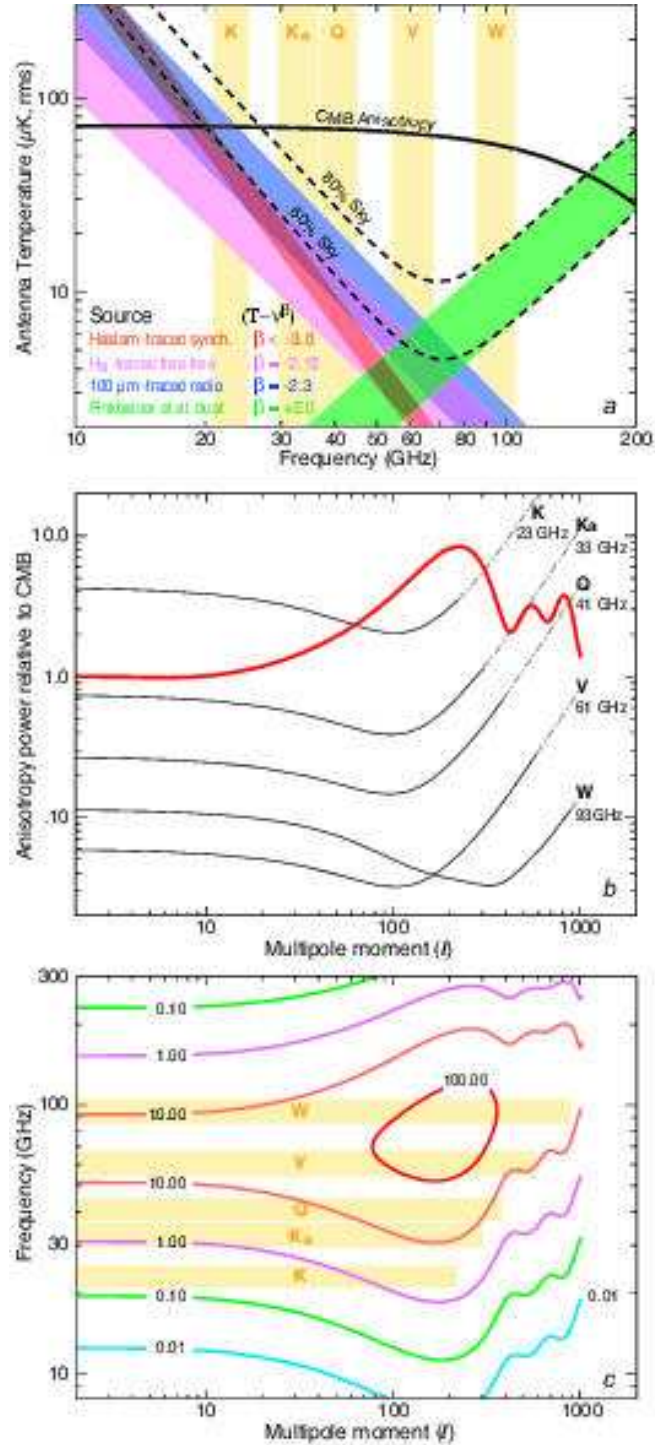}
\caption{The frequency bands were chosen so that \iMAP observes 
the CMB anisotropy in a spectral region where the emission is most dominant over the 
competing galactic and extragalactic foreground emission. (a) Spectra are shown of
CMB anisotropy (for a typical $\Lambda$CDM model) and estimates of galactic emission.
A component traced by the Haslam spatial template (red) must be steep ($\beta<-3.0$) due to 
its lack of correlation with the {\it COBE} maps.  (Note, however, that any template 
map of synchrotron emission will be frequency dependent, and hence the lack of correlation 
between Haslam and microwave maps is likely due to spatially variable spectral 
indices.)  The free-free component (pink) estimate is from H$\alpha$ data
\citep{haffner01, gaustad01}, converted assuming
$2 \mu$K R$^{-1}$ at 53 GHz and a $-2.15$ spectral index.  
A component traced by 100 $\mu$m dust emission 
(blue) has a spectral index of $\beta \approx -2.3$ \citep{kogut96b}.  This
is likely to include the flat spectrum synchrotron emission that is relatively
under-represented by 408 MHz Haslam template, but may be most of the 
synchrotron emission at microwave/millimeter wavelengths. Spinning dust emission
components should be picked up in the H$\alpha$ (pink) and 100 $\mu$m (blue) estimates.
The three component estimates, above, are partially redundant so they are added in quadrature 
to arrive at the estimate for the overall combined foreground spectra (dashed curves). 
The thermal dust emission model (green) of \citet{fink99} is a fit to {\it COBE} data.
The total galactic emission estimate is shown for cuts of the brightest microwave sky regions, 
leaving 60\% and 80\% of sky.
(b) The spatial spectra are shown, in thermodynamic temperature, relative to a typical 
$\Lambda$CDM CMB model. 20\% of the brightest galactic sky has been masked.   
The extragalactic source contribution of \citet{toffo98} is used with the assumption
that sources down to 0.1 Jy have been removed. 
(c) The contour plot shows the ratio of CMB to foreground anisotropy power as a function 
of frequency and multipole moment.  As can be seen, 
the \iMAP bands were chosen to be in the only region where the CMB anisotropy power 
is $>10\times$ to $>100\times$ that of the competing foregrounds. The \iMAP bands extend to an
$l_{\rm{max}}$ such that the beam window function is 10\%.
\label{galaxy}}
\end{figure*}

Galactic foreground signals are distinguishable from CMB anisotropy by
their differing spectral and spatial distributions. Figure \ref{galaxy}
shows the estimated spectra of the galactic foreground signals relative to the 
cosmological signal. Four physical mechanisms that 
contribute to the galactic emission are synchrotron radiation, free-free
radiation, thermal radiation from dust, and radiation from charged spinning
dust grains \citep{erickson57, fink2001, draine99, draine98a, draine98b}. 
\iMAP is designed with five frequency bands, seen in Figure \ref{galaxy},  
for the purpose of separating the CMB anisotropy from the 
foreground emission.

Microwave and other measurements show that at high galactic
latitudes ($\left| b\right| >15^\circ$) CMB anisotropy dominates the galactic 
signals in the frequency range $\thicksim 30-150$ GHz \citep{tegmark00, tegmark96}. 
However, the galactic foreground will need to be measured and removed from 
some of the \iMAP data.
There are three conceptual approaches that can be used, individually or in combination, 
to evaluate and remove the galactic foreground. 

The first approach is to use existing galactic maps at lower (radio)
and higher (far-infrared) frequencies as foreground emission templates. 
These emission patterns can be scaled to the \iMAP frequencies and subtracted.  
Uncertainties in the external data and scaling errors due to position-dependent 
spectral index
variations are the major weaknesses of this technique. There is no good microwave 
free-free
emission template because there is no frequency where it clearly dominates 
the microwave emission. High-resolution, large-scale, maps of H$\alpha $
emission \citep{dennison02, haffner01, gaustad01} can serve as a template for the free-free 
emission, except in regions of high H$\alpha$ optical depth.  
The spatial distribution of synchrotron radiation has been mapped over the
full sky with moderate sensitivity at 408 MHz \citep{haslam81}. Low frequency ($<
10$ GHz) spectral studies of the synchrotron emission indicate that the
intensities are reasonably described by a power-law with frequency 
$S\varpropto \nu^{\alpha}$ where $S(\nu)$ is the flux density, or $T\varpropto
\nu^{\alpha-2}\equiv\nu^{\beta }$ where $T(\nu)$ is the antenna temperature and $\beta \thickapprox
-2.7$, although substantial variations from this mean occur across the 
sky \citep{reich88}. There is also
evidence, based on the local cosmic ray electronic energy spectrum, that the
local synchrotron spectrum should steepen with frequency to $\beta
\thickapprox -3.1$ at microwave frequencies \citep{bennett92}.  However, this 
steepening effect competes against an effect that flattens the 
overall observed spectrum.  The steep spectral index synchrotron components seen at 
low radio frequencies become weak relative to any existing flat spectral index 
components as one scales to the higher microwave frequencies. The synchrotron 
signal is complex because individual source components can have a range of spectral indices
causing a synchrotron template map of the sky to be highly frequency-dependent.
The dust distribution has been mapped over the full sky in several infrared bands, 
most notably by the {\it COBE} and {\it IRAS} missions.  A full sky template 
is provided by \citet{schlegel98} and is extrapolated in frequency by 
\citet{fink99}.

The second approach is to form linear combinations of the multi-frequency \iMAP
sky maps such that the galactic signals with specified spectra are canceled, leaving only 
a map of the CMB.
The linear combinations of multi-frequency data make no assumptions about
the foreground signal strength, but require knowledge of the spectra of the
foregrounds.  The dependence on constant spectral indices with frequency in this
technique is less problematic than in the template technique, above, since the  
frequency range is smaller.  The other advantage of this method is that it relies only on
\iMAP data, so the systematic errors of other experiments do not enter.  The major 
drawback to this technique is that it adds significant 
extra noise to the resulting reduced galactic emission CMB map.

The third approach is to determine the spatial and/or spectral properties of 
each of the galactic emission mechanisms by performing a fit to either the \iMAP 
data alone, or in combination with external data sets. \citet{tegmark00} is an example of 
combining spectral and spatial fitting.  Various constraints can be 
used in such fits, as deemed appropriate. A drawback of this approach is the  
low signal-to-noise ratio of each of the galactic foreground components at high 
galactic latitude.  This approach also adds noise to the resulting reduced 
galactic emission CMB map.

All three of these techniques were employed with some degree of success with 
the {\it COBE} data \citep{bennett92}.  In the end, these techniques served to demonstrate 
that, independent of technique, a cut of the strongest regions of foreground emission was 
all that was needed for most cosmological analyses.  

In addition to the galactic foregrounds, extragalactic point sources will contaminate the
\iMAP anisotropy data.  Estimates of the level of 
point source contamination expected at the \iMAP frequencies have been made based on 
extrapolations from measured counts at higher and lower frequencies 
\citep{park2002, sokasian01, ref00, toffo98}.  Direct 15 GHz source count measurements 
by \citet{taylor01} indicate that these extrapolated source counts underestimate the true
counts by a factor of two.  This is because, as in the case of galactic emission discussed above, 
flatter spectrum synchrotron components increasingly dominate over steeper spectrum components
with increasing frequency.  Microwave/millimeter wave observations preferentially sample flat
spectrum sources.  Techniques that remove galactic signal 
contamination, such as the ones described above, will also generally reduce  
extragalactic contamination. For both galactic and extragalactic contamination,
the most affected \iMAP pixels should be 
masked and not used for cosmological purposes. After applying 
a point source and galactic signal minimization technique and masking the most contaminated
pixels, the residual contribution must be accounted for as a systematic error.

Hot gas in clusters of galaxies will also contaminate the maps by shifting the spectrum of the 
primary anisotropy to create a Sunyaev-Zeldovich decrement in the \iMAP frequency 
bands. This is expected to be a small effect for \iMAP and masking a modest number
of pixels at selected known cluster positions should be adequate.

Figures \ref{galaxy} illustrates how the \iMAP frequency bands were chosen to maximize the ratio 
of CMB-to-foreground anisotropy power.  After applying data cuts for the most contaminated
regions of sky, the methods discussed above are expected to substantially reduce the residual 
contamination.

\section{INSTRUMENT DESIGN}

\subsection{Overview}

The instrument consists of back-to-back Gregorian optics that feed sky signals from 
two directions into ten 4-channel polarization-sensitive 
receivers (``differencing assemblies'').  The HEMT amplifier-based receivers cover 
5 frequency bands centered from 23 to 93 GHz.  Each pair of channels is a
rapidly switched differential radiometer designed 
to cancel common-mode systematic errors.  The signals are square-law detected, 
voltage-to-frequency digitized, and then down-linked.

\subsection{Optical design}\label{optics}

The details of the \iMAP optical design, including beam patterns and sidelobe levels,
are discussed by \citet{page02}. We provide an overview here.

Two sky signals, from directions separated in azimuth by $\sim 180^\circ$ 
and in total angle by $\sim 141^\circ$, are reflected via two nearly 
identical back-to-back primary reflectors
towards two nearly identical secondary reflectors and into 20 feed horns, 10
in each optical path. The off-axis Gregorian design 
allows for a sufficient focal plane area, a compact configuration that
fits in the Delta-rocket fairing envelope, two opposite facing 
focal planes in close proximity to one another, and an unobstructed beam with 
low sidelobes. The principal focus of each optical path is between its 
primary and its secondary.

The reflector surfaces are ``shaped'' (i.e.
designed with deliberate departures from conic sections) to optimize
performance. YRS Associates of Los Angeles, CA, carried out many of the relevant 
optical optimization calculations. Each primary is a (shaped) elliptical 
section of a paraboloid with a
1.4 m semi-minor axis and a 1.6 m semi-major axis. When viewed along the
optical axis, the primary has a circular cross-section with a diameter of
1.4 m. The secondary reflectors are 0.9 m $\times$ 1.0 m. 
The reflectors are constructed of a thin carbon fiber shell over a 
Korex core, and are fixed-mounted onto a carbon-composite (XN-70 and M46-J) truss
structure. The reflectors and their supporting truss structure were 
manufactured by Programmed Composites Incorporated.  Use of the composite materials 
minimizes both mass and on-orbit
cool-down shrinkage. The reflectors are fixed-mounted so as to be in focus when
cool, so ambient pre-flight measurements are slightly out of focus. 
The reflectors have approximately 2.5 $\mu $m of vapor-deposited aluminum and 2.2 $
\mu $m of vapor-deposited silicon oxide (SiO$_x$). The silicon oxide
over the aluminum produces the required surface thermal properties 
(a solar absorptivity to thermal emissivity ratio of $\thickapprox 0.8$ with 
a thermal emissivity of 
$\thickapprox $0.5) while negligibly affecting the microwave signals.  The
microwave emissivity of coupon samples of the reflectors were measured in 
the lab to be that of bulk aluminum.  The coatings were applied by Surface 
Optics Incorporated.  

\begin{figure*}[tb]
\figurenum{6}
\plotone{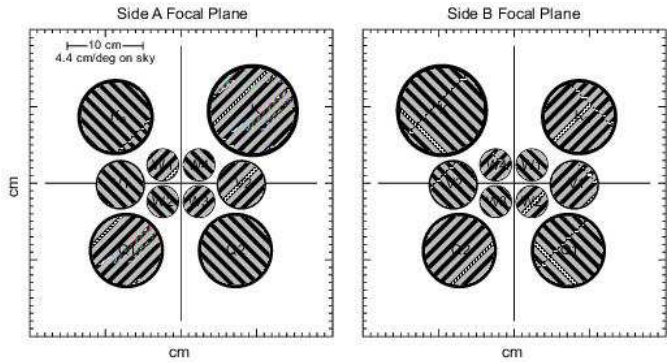}
\caption{A view of the focal plane feed layout 
in the Focal Plane Assembly (FPA) as seen from the secondaries.  The 
A-side is the $+y$ direction and the B-side is the $-y$ direction.  
The cross-hatch indicates the direction of the E-plane polarization 
for the axial OMT port.  Each radiometer is named after the 
position of its feed pair, and the OMT ports to which it attaches. 
(We sometimes refer to feeds K and Ka as K1 and
Ka1 despite the lack of a K2 and Ka2.)  
For example, for the radiometer V21, the last digit 1 corresponds to the 
axial OMT port while a last digit of 2 would indicate the radial OMT port.
Radiometer V21 differences the polarizations shown by the hatching in this 
Figure.  Radiometer V22 differences the opposite polarizations.
With this convention, the meanings 
of all of the radiometer names can be immediately found from the 
above diagram.\label{feedhorns}}
\end{figure*}
The layout and polarization directions of the 10 feeds,
covering five frequency bands, are shown in Fig. \ref{feedhorns}.
The feed designs are driven by performance requirements (sidelobe response,
beam symmetry, and emissivity), and by engineering considerations (thermal stress, 
packaging, and fabrication considerations), and by the need to assure close 
proximity of each feed tail to its differential partner. 

The feeds are designed to illuminate the primary equivalently in all bands, thus the feed 
apertures approximately scale with wavelength.  
The smallest, highest frequency feeds are placed near the center of the focal
plane where beam pattern aberrations are smallest. The HE$_{11}$ hybrid mode dominates
the corrugated feed response, giving
minimal sidelobes with high beam symmetry and low loss. The lowest frequency feeds are
profiled to minimize their length, while the highest frequency feeds are
extended beyond their nominal length so that all
feeds are roughly of the same length.  The feeds were specified and machined by 
Princeton University and designed by YRS Associates.  They 
are discussed in greater detail by \citet{barnes02}.

\subsection{Radiometer design}\label{radiometer}

The details of the radiometer design, including noise and 1/$f$ properties, are discussed 
by \citet{jarosik02}. We provide an overview here.

{\it MAP}'s ``Microwave System'' consists of ten 4-channel differencing assemblies,
each of which receives two orthogonally polarized signals from a pair of feeds. 
Each differencing assembly has both warm and cold amplifiers.
The cold portion of each
differencing assembly is mounted and passively-cooled in the 
Focal Plane Assembly (FPA) box; the warm portion is mounted in the Receiver 
Box (RXB).

\begin{figure*}
\figurenum{7}
\epsscale{2.0}
\plotone{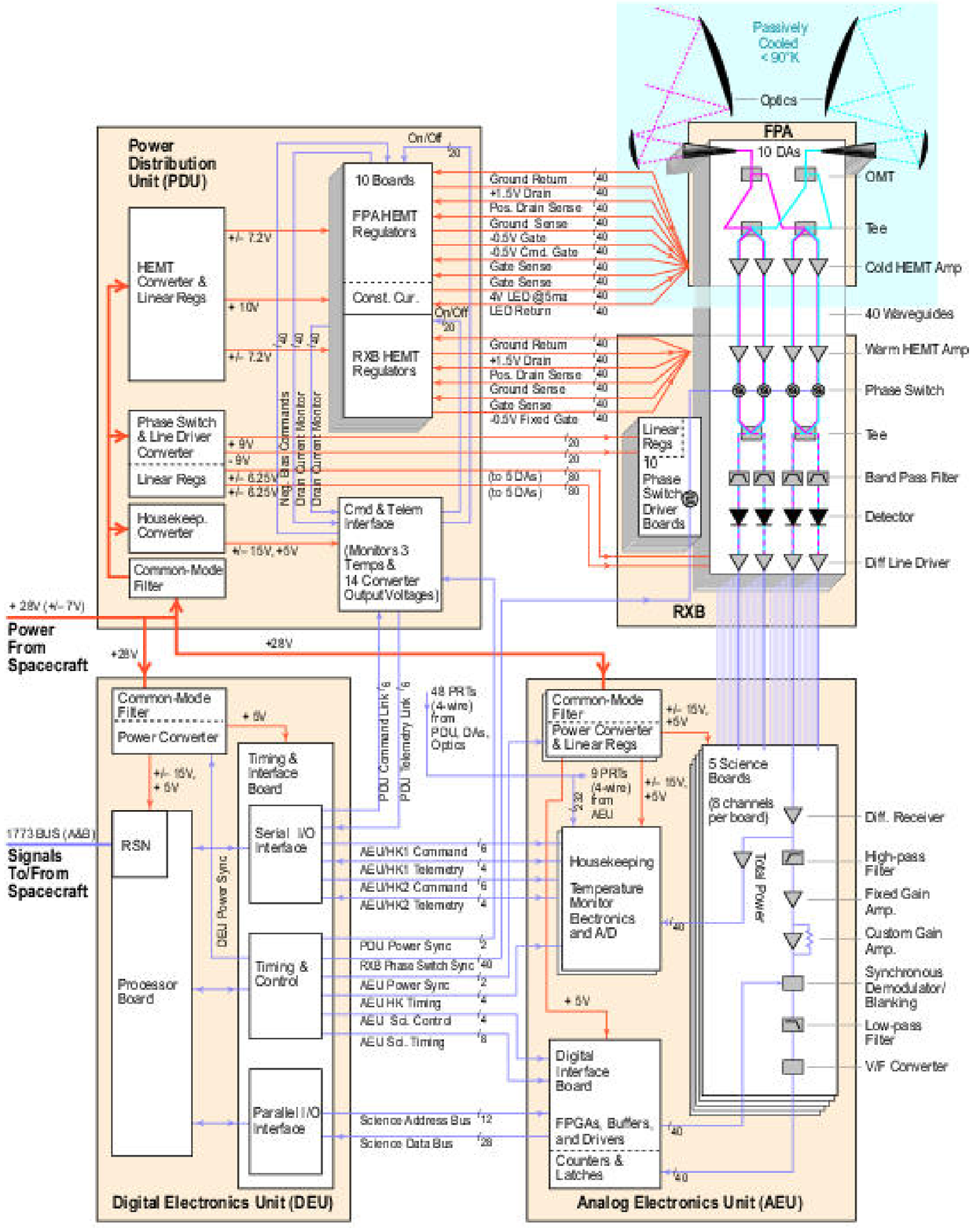}
\epsscale{1.0}
\caption{The instrument functional block diagram shows the
interconnections between the instrument electronics and the differencing
assemblies. Bold red lines are raw spacecraft power, thin red lines are
instrument power, and thin blue lines are instrument signals.\label{instblock}}
\end{figure*}
As seen in Fig. \ref{instblock}, the signal from each feed passes 
through a low-loss orthomode transducer
(OMT), which separates the signal into two orthogonal polarizations. The
$A$-side signal is differenced against the orthogonally polarized signal 
from the opposite feed, $B^\prime$. This differencing is accomplished by 
first combining $A$ and $B^\prime$ in a hybrid Tee, amplifying the two 
combined outputs in two cold HEMT amplifiers, and sending the 
outputs to the RXB via waveguide.
In the RXB the two signals are amplified in two warm HEMT amplifiers.
Then one signal path is phase-switched (0$^\circ$ or 180$^\circ$ relative to 
the other) with a 2.5 kHz square-wave modulation.  
The two signals are combined back into 
$A$ and $B^\prime$ by another hybrid Tee, filtered, square-law detected, 
amplified by two line
drivers and sent to the Analog Electronics Unit (AEU) for synchronous
demodulation and digitization. The other pair of signals, $A^\prime$ 
and $B$, are differenced in the same manner.  In \iMAP jargon, each of 
these pairs of signals comes from a ``radiometer" and both 
pairs together from a ``differencing assembly."  In all, there are
20 statistically independent signal ``channels.''

The splitting, phase-switching, and subsequent combining of the signals
enhances the instrument's performance in two ways: (1) Since both signals
to be differenced are amplified by both amplifier chains, gain fluctuations
in either amplifier chain act identically on both signals, so {\em common
mode gain fluctuations cancel}; and (2) The phase switches introduce a 
180$^\circ$ relative phase change between two signal paths, thereby interchanging
which signal is fed into each square-law detector.  Thus, {\em low
frequency (1/f) noise from the detector diodes is common mode and also
cancels}, further reducing susceptibility to systematic effects.

The first stage amplification operates at a stable low temperature to obtain
the required sensitivity.  HEMT amplifier noise decreases smoothly and only
gradually with cooling; there are no sharp break-points.  HEMT amplifiers
exhibit larger intrinsic gain fluctuations when operated cold than when
operated warm, so as many gain stages as possible operate warm, consistent with
achieving the optimal system noise temperature.

The gate voltage of the first stage of the cold HEMT amplifiers is
commandable in flight to allow amplifier performance to be optimized after
the FPA has cooled to a steady-state temperature or as the device ages.  
Each pair of
phase-matched chains (both the FPA and RXB portions) can be
individually powered off in flight to prevent any failure modes (parasitic
oscillations, excessive power dissipation, etc.) from interfering with the
operation of other differencing chains.

The frequency bands \citep{jarosik02, page02} were chosen 
to lie within commercial standards to allow
the use of off-the-shelf components.  The HEMT amplifiers 
\citep{posp00, posp94, posp92} were custom-built
for \iMAP by the National Radio Astronomy Observatory, based on custom 
designs by Marian Pospieszalski.  The HEMT devices were manufactured by 
Loi Nguyen at Hughes Research Laboratories.  The highest frequency band used  
unpassivated devices and the lower four frequency bands used
passivated devices.
The phase switches and bandpass filters were manufactured by Pacific
Millimeter, the Tees and diode detectors by Millitech, the OMTs by Gamma-f, the
thermal-break waveguide by Custom Microwave and Aerowave.  Absorber materials, 
which are used to damp potential high-Q standing waves in the box cavities of
the FPA and RXB, are from Emerson \& Cummings.  The differencing assemblies were 
assembled, tested, and characterized at Princeton University and the FPA and 
RXB were built up, aligned, tested, and characterized with their flight electronics 
at Goddard.

While noise properties were measured on the ground, the definitive noise values must be
derived in flight since they are influenced by the specific temperature
distribution within each radiometer.

The output of a square-law detector for an ideal differential radiometer 
is a voltage, $V$, per detector responsivity, $s$,
\begin{eqnarray*}
V/s & = &\left(\frac{A^{2}+B^{2}}{2}+n_{1}\right)g_{1}^{2}(t) \\
& & +\left(\frac{A^{2}+B^{2}}{2}+n_{2}\right)g_{2}^{2}(t) \\
& & \pm \frac{A^{2}-B^{2}}{2}g_{1}(t)g_{2}(t),
\end{eqnarray*}
where $g$ and $n$ are the total gain and noise of each arm of a radiometer
and $A$ and $B$ are the input voltages at the front-end of the radiometers.  
The first two terms are the total power signals.  The $\pm $ on the third term 
alternates with the 2.5 kHz phase-switch rate, with the two
arms of the radiometer always having opposite signs from one another.  The 
difference between paired detector outputs for an ideal system,
\begin{displaymath}
V/s=(A^{2}-B^{2})g_{1}(t)g_{2}(t)=(T_A-T_B)g_{1}(t)g_{2}(t)
\end{displaymath}
is used to make the sky maps.
See \citet{jarosik02} for a  
discussion of the effects of deviations from an ideal system.

\subsection{Instrument electronics design}\label{pduaeu}

There are three instrument electronic components (see Fig. \ref{instblock}).  
The Power Distribution Unit (PDU) 
provides the instrument with its required regulated and filtered power signals.
The Analog Electronics Unit (AEU) demodulates and filters the instrument detector 
outputs and converts them into digital signals.  The Digital Electronics Unit (DEU), 
built into the same aluminum housing as the AEU, holds the instrument computer and 
provides the digital interface between the spacecraft and the PDU and AEU.
The instrument electronics were built at Goddard.

The 5 science boards in the Analog Electronics Unit (AEU) take in the 40
post-detection signals from the RXB's differencing assemblies 
through differential receivers.  The 40 total power
signals are split off and sent to the AEU housekeeping card for eventual
telemetry to the ground.  The total power signals are not used in making the 
sky maps due to their higher susceptibility to potential systematic
effects, but they are useful signals for tracing the operation of the
differencing assemblies and the experiment as a whole. After the total power 
signal is split off, the remaining 
signal is sent through a high-pass filter, a fixed gain amplifier, and then
through another fixed-gain amplifier whose gain is set on the ground using
precision resistors to accommodate the particular differencing assembly
signal level.  The signal is then demodulated synchronously at the 2.5 kHz
phase-switch rate. A blanking period of 6 $\mu$sec (from 1 $\mu$sec before
through 5 $\mu$sec after the switch event) is supplied to avoid
systematic errors due to switching transients.  The $\sim40$ Hz bandwidth 
demodulated signal is
then sent through a 2-pole Bessel low-pass filter with its 3 dB point at 100
Hz.  Finally, the signal is sent through a voltage-to-frequency (V/F)
converter, whose output is latched for read-out by the processor in
the Digital Electronics Unit (DEU), before being losslessly compressed
and telemetered to the ground.  The AEU has a digital interface board and 
two power converter boards, which supply
the requisite $\pm 15$ V, $\pm 12$ V, and $+5$ V to the other AEU boards.

The noise in each of the 40 AEU signal channels is limited to $<150$ nV Hz$^{-0.5}$ 
from 2.5 to 100 kHz to ensure that the AEU contributes $<1$\% of the
total radiometer noise. The AEU channel bandwidth is 100 kHz. The gain
instability is $<5$ ppm for synchronous variations with the Observatory spin.
This requires that the components be thermally stable to $<10$ mK at the
Observatory spin rate. Random gain instabilities are limited to 
$<100$ ppm from 8 mHz (the spin frequency)
to 50 Hz. The DC-coupled amplifier has random offset variations $<1$ mV rms 
from 8 mHz to 50 Hz to limit its contribution to the post-demodulation noise. 

The AEU also contains two boards for handling voltage and high precision 
temperature sensing of the instrument. These monitors go well beyond the
usual health and safety functions; they exist primarily to confirm 
voltage and thermal stability of the critical items in the instrument.  
Should variations be seen, these monitor signals can be used as tracers
and diagnostics to characterize the effects on the science signals.

The DEU receives power from the spacecraft bus (fed through the PDU), 
applies a common-mode filter, and uses a DC-DC power converter to 
generate +5V and
$\pm$15 V for internal use on its timing and interface board and on its
processor board.  The power converter is on one 
card while the Remote Services Node (RSN) (see \S \ref{rsn}) and timing boards 
are on opposite sides of a double-sided card.

\begin{figure*}
\figurenum{8}
\plotone{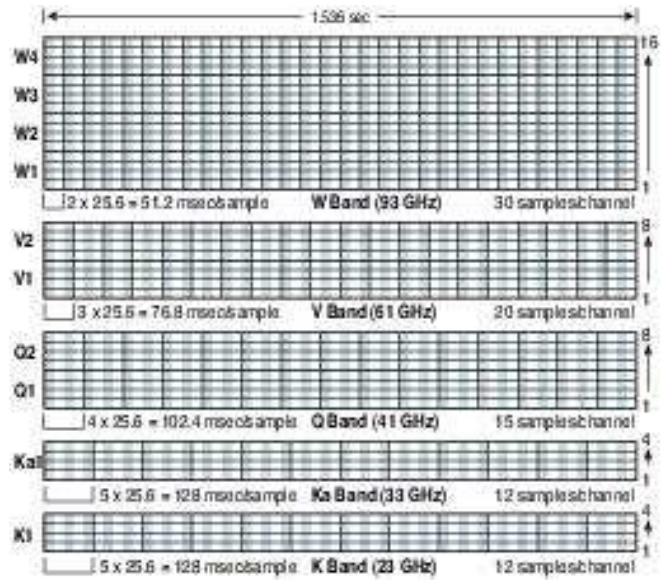}
\caption{All 40 channels from the ten differencing assemblies 
are sampled in multiples of an underlying 25.6 ms period.  The number of 25.6 ms 
periods that make up each sample is chosen with regard to the beamwidth of that
channel to avoid smearing a source on the sky.  Every 1.536 s the samples are 
collected and put into packets for later data down-link. \label{samples}}
\end{figure*}
The DEU provides a 1 MHz ($\pm 0.005$\% , 50\% duty cycle) clock, 
derived from a 24 MHz crystal oscillator, to the V/F converter in the 
AEU.  
The DEU also supplies a 100 kHz clock to the power converters in the 
PDU and the AEU, a 2.5 kHz (50\% duty cycle) clock to the RXB and AEU for
phase switching, and a 5 kHz pulse to the AEU for blanking the science
signal for 6 $\mu$sec during the 2.5 kHz switch transitions.  The DEU 
also provides a 25.6 msec (64 cycles at 2.5 kHz = 39.0625 Hz) 1 $\mu$sec
wide negative logic clock to the AEU for latching the 14-bit 
science data samples.  All 40 channels are integrated in the AEU and sent 
to the DEU every 25.6 msec, as shown in Figure \ref{samples}.
The RF bias (total power) signals from the 40 AEU 
channels and 57 platinum resistance thermometer (PRT) temperature 
signals are passed from the AEU to the DEU every 23.04 seconds.
All of these DEU clock signals are synchronous with the 24 MHz master clock.  
The DEU
also sends voltage, current, and internal temperature data from the
PDU, AEU, and DEU.

The 69R000 processor in the DEU communicates with the main computer 
(see \S{\ref{cdh}}) over a 1773 optical fiber bus.  
The DEU uses 12k (16-bit 
words) memory for generic RSN instruction code, 24k for DEU-specific code, 
and 10k for data storage.

The AEU and DEU are packaged together in an aluminum box enclosure with 
shielding between the AEU and DEU sections. The AEU/DEU and the PDU 
dissipate 90-95\% of their power from their top radiators. 
They are qualified over a $-10$ to $+50$ C temperature range, but normally
operate over a $0-40$ C range.  The temperature variations of the boxes
are designed to be limited to $<10$ mK peak-to-peak at the spin period.
The 100 mil effective box wall thicknesses allow the electronics to survive the
space radiation environment (see \S\ref{radiation}).
 
The PDU receives 21-35 V from the spacecraft, with spin-synchronous variations
$<0.5$ V peak-to-peak, and provides all instrument power.  Every HEMT gate and drain is
regulated with a remote-sense feedback circuit.  The PDU clamps the voltage 
between the gate and drain to be $<2.1$ V (at 10 $\mu$A) to prevent damage 
to the sensitive HEMT devices.
The drain voltages are commandable in 8 steps ($\sim 70$ mV resolution)
from 1.0 to 1.5 V, and the gates are commandable in 16 steps ($\sim 35$ 
mV resolution) from $-0.5$ V to 0 V.  Voltage drifts are $\ll 10$ mV for
the drain, and $\ll 5$ mV for the gates over the 0-40 C operating range.

The broadband noise requirements on the HEMT gate and drain supplies are:
$<23f^{-0.45}$ $\mu$V Hz$^{-0.5}$ for 0.3 mHz - 1 Hz 
                              ($=885\;\mu$V Hz$^{-0.5}$ at 0.3 mHz); 
$<23\; \mu$V Hz$^{-0.5}$ for 1-50 Hz; and
$<100$ nV Hz$^{-0.5}$ for 2.5 kHz and its harmonics to 50 kHz.  
These frequency ranges correspond to
the precession frequency ($\sim 0.3$ mHz), the spin frequency ($\sim 7.57$ mHz),
and the phase-switch frequency ($\sim 2.5$ kHz), respectively.
Spin-synchronous rms variations are $<400$ nV.

The PDU supplies 4 V at 5 mA ($<5$ nA rms variation at the spin
period) to two series LEDs on each cold HEMT amplifier.  The LED light 
helps to stabilize the gain of the HEMT devices.

The PDU supplies $\pm 9$V ($<50$ mV ripple, $<20$ mV common mode noise) 
to the phase switch driver cards, which are mounted in the RXB, near the 
differencing assemblies.  The PDU also supplies $\pm 6.25$ V to the 
line drivers in the RXB.

The PDU allows for on/off commands to remove power from any one or more
of the 20 radiometers (phase-matched halves of the differencing assemblies).
Should several radiometers be turned off such that the lack of power dissipation drives
the PDU temperature out of its operational temperature range, a 
supplemental make-up heater can be commanded on to warm the PDU.

\subsection{Instrument Calibration}

The instrument is calibrated in-flight using observations of the CMB dipole and of Jupiter, 
as discussed in \S \ref{syserr}.

Despite the in-flight amplitude calibration (telemetry digital units per 
unit antenna temperature), it was necessary to 
provide provisional calibration on the ground to assess and characterize various aspects 
of the instrument to assure that all requirements would be met.  For example, cryogenic microwave 
calibration targets (``x-cals") were designed and built to provide a known temperature for each 
feed horn input for most ground tests.  The x-cals, attached directly to each of the 20 feed 
apertures, were individually temperature-controlled to a specified temperature in the range of 15 K
to 300 K, and provided for temperature read-out. 

Observations of Jupiter and other celestial sources provide an in-flight pointing offset 
check relative to the star tracker pointing.  The pointing directions of the feeds were 
measured on the ground 
using standard optical alignment techniques.  Jupiter also serves as the source for beam 
pattern measurements in-flight.  Beam patterns were measured in an indoor compact antenna 
range at the Goddard Space Flight Center and the far-sidelobes were measured between rooftops 
at Princeton University.  

\section{OBSERVATORY DESIGN}

\begin{figure*}
\figurenum{9}
\epsscale{2.0}
\plotone{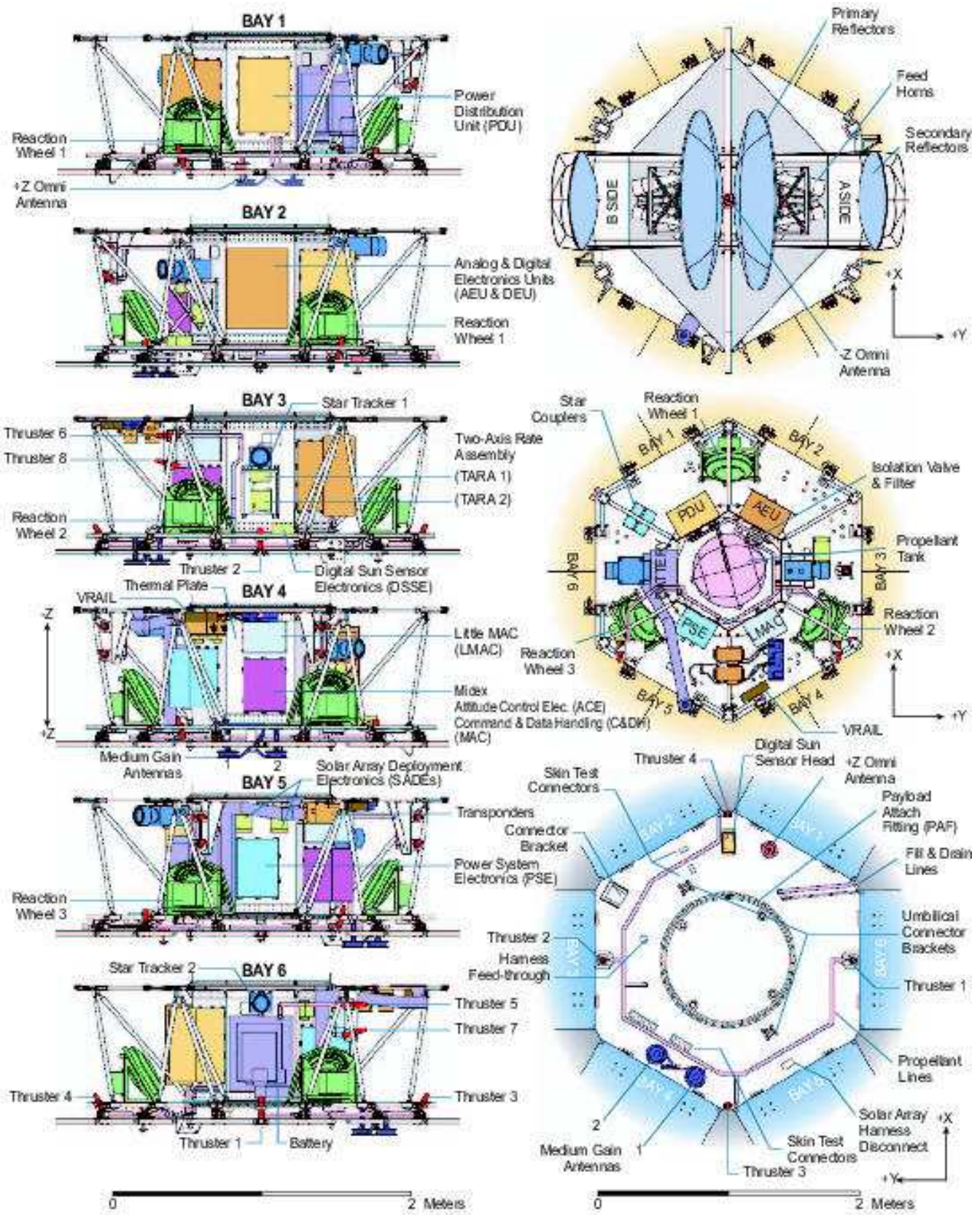}
\epsscale{1.0}
\caption{The physical lay-out of the Observatory is shown 
from various perspectives.\label{sclayout}}
\end{figure*}

\begin{figure*}
\figurenum{10}
\epsscale{2.0}
\plotone{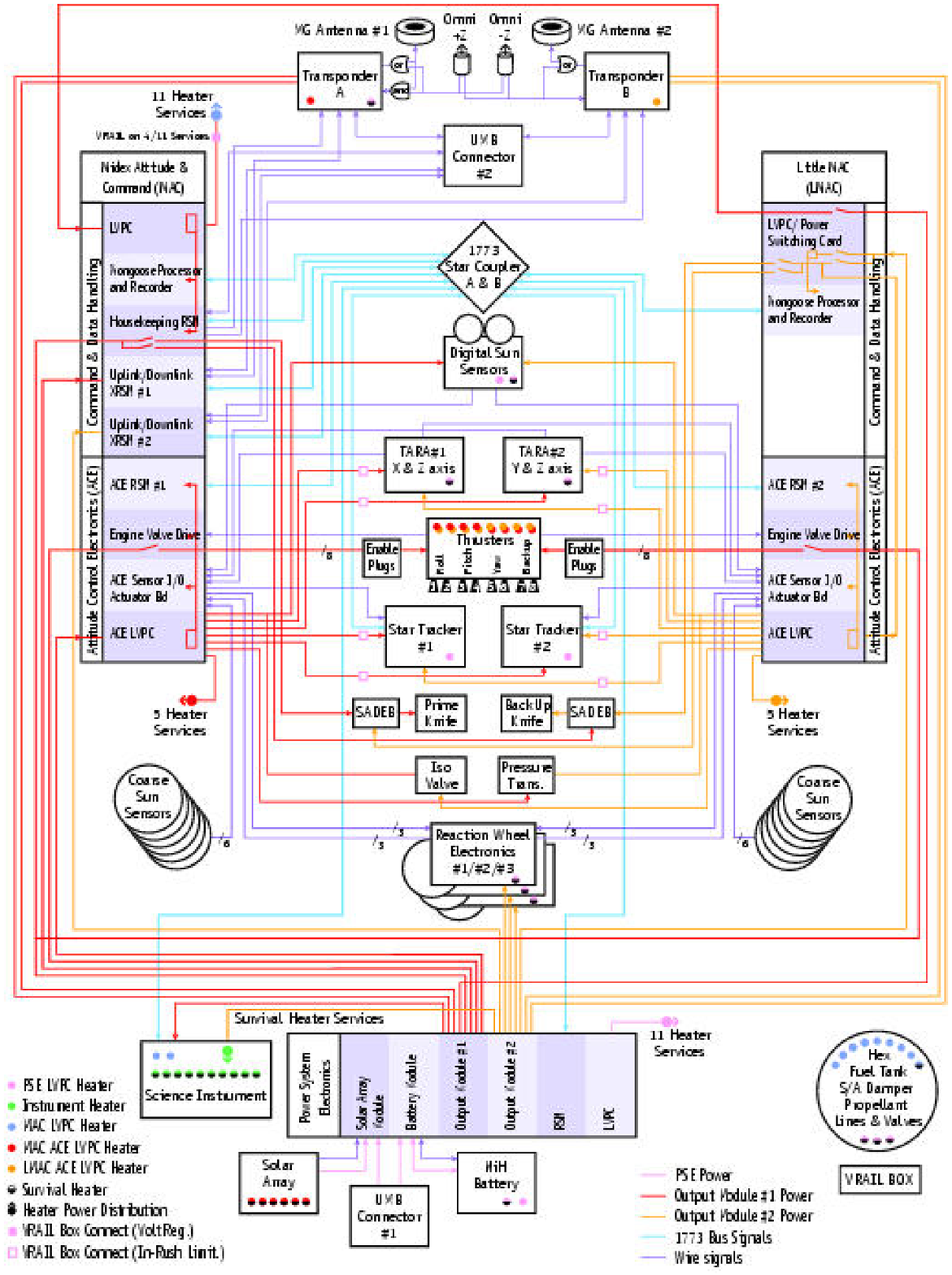}
\epsscale{1.0}
\caption{The functional block diagram of \iMAP shows the
interconnections between the components.\label{scblock}}
\end{figure*}

This section provides an overview of the \iMAP spacecraft systems. The physical layout of 
the Observatory is shown in Fig. \ref{sclayout} and the functional
block diagram is shown in Fig. \ref{scblock}. 

\subsection{Command and Data Handling\label{cdh} \label{rsn}}

\iMAP implements a distributed architecture with a central Mongoose computer,
which communicates via a 1773 (Spectran 171.2 $\mu $m 1300 nm polyamide) fiber 
optic bus to Remote Services Nodes (RSNs) located in the electronics boxes 
(see Figure \ref{scblock}).  Each RSN provides a standardized interface 
to analog and digital electronics and uses common flight software. 
The RSN circuitry
occupies half of one side of a printed circuit card; the remainder of the
card space is used for application-specific circuitry.
The fiber optics are interconnected using two (redundant) star couplers. 
Each is a pigtail
coupler with a $16\times 16$ configuration in a 8.9 cm x 20 cm x 4.4 cm 
aluminum housing.

Attitude Control Electronics (ACE) and Command and Data Handling (C\&DH) functions 
are housed in the Midex ACE and C\&DH (MAC) box, which contains 9 boards: 
(1) A Mongoose V R3000 32-bit rad-hard RISC processor board 
includes 4 MB of EEPROM memory and 320 MB of DRAM  
memory (of which 224 MB is for the solid state data ``recorder'', 32 MB is for
code, and 64 MB is for check-bytes), 
with 4 Mbps serial outputs to the transponder interface boards and redundant
1773 interfaces;
(2,3) Two Low Voltage Power Converter (LVPC) boards (see \S \ref{power});
(4,5) Two up- and down-link transponder boards, which are both always active
(see \S \ref{rfcomm});
(6) A housekeeping board, which monitors 6 deployment potentiometers (see \S \ref{deploy}), 
9 status indicators,
46 temperature channels, and 2 voltages, in addition to one spare word, for a total
of 64 input signals;
(7) An ACE RSN and sensor electronics I/O board, which reads the digital and coarse sun 
sensors, and reads and commands the reaction wheels (see \ref{acsrcs});
(8) An ACE sensor I/O board that queries and/or reads information from the 
inertial reference unit (gyro), digital and analog Sun sensors, 
reaction wheels, separation switches (see \ref{acsrcs}),  
and sends a timing pulse to the star trackers and monitors thermistors and 
solar array potentiometers; and
(9) A propulsion engine valve drive (EVD) electronics board that controls 
the 8 thrusters (see \S \ref{acsrcs}).

Selected redundancy is provided by a ``Little MAC'', or LMAC box.  The LMAC
box houses a total of six boards.  Four boards hold the redundant attitude control 
electronics (an LVPC board,
a EVD electronics board, a sensor I/O board, and an RSN board).  The two remaining 
boards are 
a redundant Mongoose processor board, and a LVPC board with power switching circuitry that
controls the redundancy between the MAC and the LMAC functions.  Only one
Mongoose processor (MAC or LMAC) is on and in control at any one time. Shortly after launch
the LMAC ACE takes primary control, with the MAC ACE powered on as a ``hot'' back-up.
The active Mongoose processor board sends an ``I'm OK'' signal to the LMAC ACE.  If the 
LMAC ACE fails to get the ``I'm OK'' signal, then it places the Observatory into 
safe-hold.  That is, the ACE RSN acts as an attitude control safehold processor. 
If the housekeeping RSN fails, the LMAC ACE is powered on by ground command.  Only a 
single uplink path can be active, selected by ground command.

\subsubsection{Data Sampling and Rates}
The MAC Mongoose processor gathers the Observatory science and engineering data and arranges 
them in packets for down-link. The instrument's compressed science data comprises 53\% of the 
total telemetry volume. 

As shown in Fig. \ref{samples}, during a 1.536 sec period the instrument collects 
30 samples in each of 16 W-band (93 GHz) channels,
20 samples in each of  8 V-band (61 GHz) channels, 
15 samples in each of  8 Q-band (41 GHz) channels, and
12 samples in each of 4 Ka-band (33 GHz) and 4 K-band (23 GHz) channels.
In this way the bands with smaller beams are sampled more often. With 
856 samples all together at 16 bits per sample in 1.536 sec, there is a 
total instrument science data rate of 8917 bits \persec.
The instrument science data is put into two packets, with all of the W-band 
data in one packet and all other data in a second packet.  The second packet 
is assigned some additional filler to make the two packet lengths
identical.  Each packet also has 125 bits \persec of ``packaging'' overhead.
The adjusted total raw instrument data rate is 9042 bits \persec.  These 
data flow from the DEU to the MAC.  

The Mongoose V processor losslessly compresses this data on-the-fly by a 
factor of $\approx 2.5$ using the Yeh-Rice algorithm, and then 
records 3617 bits \persec of science data.  An additional 500 bits \persec 
of instrument house-keeping data and 2750 bits \persec of spacecraft data 
are recorded, for a total of 6867 bits \persec.  The $\approx 0.6$ Gbits 
of data per day are down-linked daily.  Of the overall down-link rate of 666666 
bits \persec, 32000 bits \persec are dedicated to real-time telemetry, and 
563167 bits \persec are allocated to the playback of the stored 
data, which takes 17.6 min.  The bit rate is commandable and may be adjusted
in flight depending on the link margins actually achieved.

\subsubsection{Timing control}
The Mongoose board in the MAC/LMAC maintains time with a 32-bit second 
counter and a 22-bit microsecond counter.  There is also a watchdog timer 
and a 16-bit external timer. The clock is available to components on the bus
with a relative accuracy of 1 ms.  Data are time-tagged so that a relative
accuracy of 1.7 ms can be achieved between the star tracker(s), gyro, and the
instrument.  The Observatory time is correlated to ground time to within 1 s.

\subsection{Attitude and Reaction Control}{\label{acsrcs}}

The attitude control system (ACS) takes over control of the orientation of the 
Observatory
after its release from the Delta vehicle's third stage. From the
post-separation $3\sigma$ initial conditions ($\pm 2^\circ$ s$^{-1}$ $x$- and $y$-axis 
tip-off rates, and $\pm 2$ rpm $z$-axis de-spin rate), \iMAP is designed to
achieve a power-positive attitude (solar array normal vector within 25$^\circ$ 
of the Sun direction) 
within 37 min using only its wheels for up to 2$\sigma $
tip-off rates.

The attitude control and reaction control (propulsion) systems  
bring \iMAP through the Earth-Sun phasing loops (see Fig. \ref{trajectory}) 
such that the thrust is within $\pm 5^\circ$ 
of the desired velocity vector with 1$^\circ$ maneuver accuracy.  After 
the lunar-swingby, \iMAP cruises, with only slight trajectory-correction 
mid-course maneuvers, into an orbit about L$_{2}$. Once there, the
ACS provides a combined spin and precession such that
the Observatory spin ($z$) axis remains at $22.5^\circ\pm 0.25^\circ$ from the 
Sun vector for all science observations.  This is referred to as ``Observing Mode.''
For all maneuvers that interrupt Observing Mode (after the mid-course correction 
on the way to L$_2$, and for station-keeping maneuvers at L$_2$) the spin ($z$) axis 
must always remain at $19\dg\pm 5\dg$ relative to the anti-Sun vector to prevent 
significant thermal changes.
The spin rate must be an order of magnitude higher than 
the precession rate and the instrument boresight scan rate must
be $2.59\dg$ s$^{-1}$ to $2.66\dg$ s$^{-1}$. The $2.784\dg$ s$^{-1}$ 
($0.464$ rpm)
spin and the $0.100\dg$ s$^{-1}$ (0.017 rpm, 1 rev hr$^{-1}$) precession 
rates are in opposite directions and are 
controlled to within 6\%. The ACS must also manage 
momentum and provide an autonomous safe-hold. Momentum management 
occurs throughout the mission, with each momentum unload leaving 
$\lsim 0.3$ Nms per axis.

The instrument pointing knowledge requirement of 1.3 arcmin ($1\sigma$) is 
sufficient for the aspect reconstruction needed to place the 
instrument observations on the sky.  Of this, 0.9 arcmin (a root-sum-square 
for 3 axes) is allocated to the attitude control system. 

The radius of the Lissajous orbit about L$_{2}$ (see Fig. \ref{trajectory}) 
must be  $\gsim 0.5\dg$ (between the Sun-Earth vector and the Earth-\iMAP vector) 
to avoid eclipses, and $<10.5\dg$ to maintain the antenna angles necessary 
for a sufficient communication link margin.

The attitude and reaction control systems include attitude control 
electronics (ACE boards, in both the MAC and LMAC boxes), 3 reaction wheels, 
2 digital Sun sensors, 6 prime plus 6 redundant coarse Sun sensors, 
1 gyro (mechanical dynamically-tuned, consisting of 2 
TARAs = Two-Axis Rate Assemblies), and 2 star trackers (one prime and one
redundant).  The propulsion 
system consists of two engine valve drive cards (one in the MAC box and a
redundant card in the LMAC box), a hydrazine propulsion tank with stainless
steel lines to 8 thrusters (2 roll, 2 pitch, 2 yaw, and 2 backups), with
an isolation valve and filter.  (The fuel filter is, appropriately 
enough, a hand-me-down from the {\it COBE} mission.)

The Lockheed-Martin model AST 201 star trackers are oriented in 
opposing directions on the $y$-axis.  They are supplied with a 1773
interface, track at a rate of 
3$^\circ$ s$^{-1}$, and provide quaternions with an accuracy of 
2.3 arcsec in pitch
and yaw, and 21 arcsec (peak) in roll. 

The TARAs are provided by Kearfott. One TARA
senses $x$ and $z$ rates and the other senses $y$ and $z$ rates over a 
12$^\circ$ s$^{-1}$ range. The TARAs provide a digital pulse train 
(as well as
analog housekeeping) with 1 arcsecond per pulse. The linear range is 
$\pm 5^\circ$ s$^{-1}$ with an angle random walk of $<0.03^\circ$ hr$^{-1}$.

The digital Sun sensors are provided by Adcole. The two digital Sun sensors
each provide a field-of-view that extends $\pm 32^\circ$ from its boresight, 
and they are positioned to provide a slight field-of-view
overlap. The output has two serial digital words and analog housekeeping.
The resolution is $0.004^\circ$\ (0.24 arc min) and the accuracy is 
0.017$^\circ$.  The 12 coarse Sun sensors (6 primary and 6 redundant) are 
also provided by Adcole. They are mounted at the outer edge of each solar panel and are 
positioned with boresight angles 
pointed alternately 36.9$^\circ$ towards the instrument (bays 2, 4, 6) and 
36.9$^\circ$ towards the Sun (bays 1,3 5) with respect to the plane of the solar arrays.
Their fields of view extend $\pm 85^\circ$ from each of the boresights.  They provide an
attitude accuracy of better than  3$^\circ$ when uncontaminated by
Earth albedo effects.  Their output is a photoelectric current.

The Ithaco Type-E reaction wheel assemblies have a maximum torque of $\pm
0.35$ Nm, but the \iMAP ACE limits this to $\pm 0.177$ Nm ($<125$ Watts per 
wheel, $<5.3$ amps per wheel) to control power use, especially for
the high wheel rates that potentially could be encountered during initial acquisition.  
{\it MAP}'s maximum momentum storage is $\pm 75$ Nms.  The wheels take an
analog torque input and provide a tachometer output.

The propulsion tank, made by PSI, was a prototype spare unit from the {\it TOMS-EP} 
Program, which kindly provided it for use on \iMAP.  It has a titanium 
exterior, which is only 0.076 cm thick at its thinnest point,
with an interior elastomeric diaphragm to provide positive expulsion of the 
hydrazine fuel.  The tank is roughly 56 cm in diameter, has a mass of 6.6 kg, 
and holds 72 kg of fuel.

The eight 4.45 N thrusters are provided by Primex.  Each thruster has a catalyst
bed heater that must be given at least 0.5 hr to heat up to at least 125 C
before a thruster is fired.  The fuel must be maintained between 10 C and 50 C
in the lines, tank, and other propulsion system components, without the use of any actively 
controlled heaters.  A zone heating system was developed to accomplish this 
relatively uniform heating of the low thermal conductivity (stainless steel) 
fuel lines that run throughout the Observatory.  The lines, which are wrapped in 
a complex manner that includes heaters and thermostats, are divided into 
thermal zones.  The zones were balanced relative to one another during Observatory 
thermal balance/thermal vacuum testing.  In flight, zone heaters can be
switched on and off, by ground command, to provide re-balancing, but this is not expected 
and, if needed, would be far 
less frequent switching than would be the case with an autonomously active 
thermal control system.  In this way the Observatory thermal and electrical 
transitions will be minimal.

A plume analysis was performed to determine the amount of hydrazine byproduct contamination
that would be deposited in key locations, such as on the optical surfaces.  These analyses 
take into account the position and angles of the eight thrusters.  The final thruster placement 
incorporated the plume analysis results so that byproduct contamination levels are 
acceptably low.

The ACS provides six operational modes, described below.

{\it Sun Acquisition} mode uses reaction wheels to orient the spacecraft
along the solar vector following the Delta rocket's yo-yo despin (to $<2$ rpm) 
and release of the Observatory from the third stage. This must be accomplished $<37$ mins 
after separation due to battery power limitations. The body
momentum is transferred to the reaction wheels until the angular rate is
sufficiently reduced, and then position errors from the coarse Sun sensors and
rate errors from the inertial reference unit are nulled. 

{\it Inertial} mode orients and holds the Observatory at a fixed angle relative 
to the Sun vector in an inertially fixed power-positive
orientation, and provides a means for slewing the Observatory between two
different inertially fixed orientations. Reaction wheels generate the motion
and the IRU (gyro) provides sensing. Inertial mode can
be thought of as a ``staging'' mode between Observing, Delta-H, or Delta-V
modes.  Information from the digital Sun sensor and star tracker are used in a
Kalman filter to update the gyro bias and quaternion error estimates and
these data are used by the controller.

{\it Observing} mode moves the Observatory in a compound spin (composed of 
a spin about the $z$-axis combined with a precession of the $z$-axis about the
anti-Sun vector) that satisfies the scientific requirements for sky scanning. The
total reaction wheel momentum is canceled by the prescribed body momentum.  The 
Kalman filter is used in the same manner as in the Inertial mode, described above.

{\it Delta-H} mode is used to change the Observatory's angular momentum. It
is used following the yo-yo de-spin from the Delta rocket's third stage for
$>2\sigma$ tip-off rates, 
and to reduce wheel momentum that accumulates later in the mission. Thrusters 
can be used as a back-up in the unexpected event that the Observatory has more 
momentum than can be handled by the
reaction wheels. A pulse width modulator is used to convert rate controller
information to thruster firing commands. The reaction
wheel tachometers are used along with the IRU to 
estimate the total system momentum.

{\it Delta-V} mode is used to change the Observatory's velocity. It
is used in the initial phasing loop burns and for station-keeping near L$_2$. 
The controller must account not only for the desired combination of thrusters 
and degree of
thruster firings, but must also assure that position and rate errors (which
may arise from center-of-mass offsets, thruster misalignments, or plume impingement) 
are maintained within allowable limits.

{\it Safe-hold} mode slews the Observatory to an inertially fixed
power-positive orientation. Reaction wheels are used to move the Observatory
according to coarse Sun sensor information. Safe-hold functions are similar
to Sun Acquisition functions, except that Safe-hold mode can use only the coarse Sun 
sensors, not the IRU.  The IRU mode allows control with higher system momentum.

\subsection{RF Communications} {\label{rfcomm}}

\iMAP carries identical prime and redundant transponders, each capable of 
communications on both
the NASA Space Network (SN) and Ground Network (GN). Each S-band (2 GHz) 
transponder has
an output power of $\geq 5\ $W. The prime and redundant transponders are
controlled by RSN boards in the MAC and LMAC boxes.

Two omnidirectional (``omni'') antennas, each with 0 dBi gain over $\pm 80^\circ$ from the 
boresight, combine to provide nearly full spherical coverage for use should the 
Observatory attitude be out of control. The omni antennas (like those on 
the XTE, TRMM, and TRACE missions) are crossed bow-tie dipoles, 
built in-house at Goddard. Because one of \iMAP's omni antennas will 
never be in sunlight during nominal science operations, the omni antenna 
and its Gore coaxial cable were tested at Goddard for 
$-200^\circ$ C operation.

\iMAP carries two medium gain antennas for high speed data transmission to Earth. 
The medium gain
antennas, designed, built and tested at Goddard, use a circular PC-board
pattern to provide $\geq 5.5$ dBi gain over $\pm 35^\circ$ from the 
boresight.   They have been designed and tested for $\pm 100^\circ$ C, 
while nominally in 100\% sunlight during most of the mission.  The
antennas have a Goretex thermal protective cover.

The RF communications scheme is shown as part of Fig. \ref{scblock}.
Each of the two transponders receives from its own medium gain and 
both omni antennas at all times.  Microwave switches allow each 
transponder to transmit to both of the omni antennas or, alternately, 
to its medium gain antenna.

\subsection{Power System} {\label{power}}

The power system provides at least 430 Watts for 30 min to assure 
that safehold can be reached, and 430 Watts of average orbital power at the 
end of 27 months for the L$_2$ orbit at $22.5\dg \pm 0.25\dg$.  The system is
designed such that the battery depth of discharge is never worse than 60\%.

The power system consists of six GaAs/Ge solar array panels, 
a NiH$_2$ battery, a Power Supply Electronics (PSE) box, and Low Voltage Power 
Converter (LVPC) cards located in other electronics boxes.

The solar panels are supplied by Tecstar, Inc. To 
keep the instrument passively cool, \iMAP requires that the backs of the 
solar panels be covered with thermal
blankets. In this unusual configuration, steps must be taken to keep the
panels from overheating. Thus, much of the Sun-side surface area that is
not covered by solar cells is covered by second surface mirrors (optical
surface reflectors) that allow the panels to reject heat by reflecting sunlight
rather than absorbing it. The solar panels are specified to supply $>678$ Watts 
at $>34.6$ Volts at the beginning of
life, and $>502$ Watts at $>31.5$ Volts after 27
months of flight, both at 86 C.  Each of the 6 solar panels includes 14
strings of 48 (4.99 cm x 5.65 cm) GaAs/Ge cells each, with a total solar cell area
of 5187 cm$^{2}$ for each of the 6 panels, with a solar absorptance of 
$>0.87$ averaged over the solar spectrum.

The nickel hydrogen (NiH$_2$) battery is a 23 amp-hr common pressure vessel
design that consists of 11 modules, each of which contains two cells
that share common electrolyte and hydrogen gas.  It is supplied by 
Eagle Picher Technology of Joplin, MO. 
The battery is capable of supplying about 35 V when freshly charged.  (The
Observatory is nominally expected to operate at about 31.5 V.) The solar 
array is deployed after Observatory separation from the 
launch vehicle, allowing the battery to recharge once the power system is 
power-positive on the Sun.  

The PSE and LVPCs were designed and built by Goddard. The PSE's LVPC supplies
switched and unswitched secondary power ($\pm15$V, $+5$V), and it provides
unregulated switched primary power to loads in the subsystem in which
it is located.

In case of emergency, the on-board computer will autonomously begin taking action, 
including shedding loads.
The power system uses a pre-programmed relationship between the battery 
temperature and voltage (a ``V/T curve'') in an active control loop.  When the battery 
reaches a 90\% state-of-charge (SOC) and the bus voltage drops below 30 V, a warning 
is issued by the Observatory and the PSE is set to its normal operating V/T relationship. 
At an 80\% state-of-charge and a bus drop to 27 V, the instrument, catalyst bed heaters, 
makeup heaters and battery operational heater are all autonomously turned off and the 
Observatory goes into safe-hold.  At 60\% state-of-charge and $<26$ V the second star 
tracker is turned off.  At 50\% state-of-charge and $<25$ V the PDU make-up heater, RXB 
operational heater, and solar array dampers heaters are all turned off.
Although all components have been tested to operate properly at as low as 21 V, this
condition can not be supported in steady-state for more than a couple of
hours.  This is long enough only for emergency procedures to take control and
attempt to fix a problem.

\subsection{Solar Array Deployment System}{\label{deploy}}

Web thermal blanket segments run from center-line to center-line on the anti-Sun 
side of the six solar array panels.  The panels are folded for launch 
to $5\dg$ from the $z$-axis, with the blankets tucked gently inside.  
The panels are held in place by a Kevlar rope that runs around the 
external circumference of the spacecraft, in V-groove 
brackets, and is tensioned to almost 1800 N against a spring-loaded hot
knife.  The knife, a circuit board with high ohmic loss traces, is
activated with a voltage that is applied based upon an on-board timing 
sequence keyed to separation of \iMAP from the launch vehicle.  Redundant 
thermal knives are on opposite sides (bays 3 and 6) of the Observatory.
An automated sequence fires one knife first, and then the second knife is energized 
after a brief time delay.

Each solar array panel is mounted with springs and dampers.  Once the
tensioned Kevlar cable is cut it departs the Observatory at high speed.  Then the panels 
and web blankets gently unfold together (in $\sim 15$ s) into the $z=0$ plane.

\subsection{Mitigation of Space Environment Risks}

\subsubsection{Charging}

Spacecraft charging can be considered in two broad categories: surface
charging and internal charging. 

The various external surfaces,
whether dielectric or conductive, will be exposed to a current of charged
particles. If various surfaces are not reasonably conductive 
($<10^9 \Omega $ square$^{-1}$) and tied to 
the Observatory ground then differential
charging will occur.  Differential charging leads to potential
differences that can discharge either in sudden and large sparking events,
or in a series of smaller sparks.  In either case these discharges can
cause severe damage.  

For space missions in low Earth orbit the local
plasma can be effective in safely shorting out potential charge build-ups. 
Components in sunlight also have the advantage of discharge via the
photoelectric effect.  \iMAP does not have the benefit of a local, high-density,
plasma, and
much of the Observatory is in constant shadow.  Thus, with only a few
well-considered exceptions, \iMAP was built with external surfaces that
are in reasonable conductive contact with one another.  Conductive surfaces 
used on \iMAP include indium-tin-oxide coatings on teflon, silver teflon, and 
paint, and carbon-loaded (``black'') kapton.

The Observatory is also exposed to a current of higher energy particles that 
penetrate the skin layer materials and can deposit charge anywhere in the 
Observatory, not just on its surface. Only
radiation shielding in the form of mass can stop the high energy particles,
and mass is always a precious resource for a space mission.  \iMAP used a
complex set of implementation criteria to protect against internal 
charging. This protection can not be absolute: 
there is a power spectrum of incoming radiation,
the current amplitudes as a function of location are uncertain, 
and the stopping power of shielding is statistical. 
In general, the \iMAP interior surfaces are
grounded where possible, and a minimal amount of external shielding metal (either 
boxes, plates, or added lead shielding) is used to reduce the size of the charging 
currents.  All susceptible circuits were shielded with an equivalent stopping
power of 0.16 cm of aluminum.  For example, 0.2 mm of lead foil was wrapped 
around all the bias and control lines of the HEMT amplifiers to augment the existing 
harness shielding.

\subsubsection{Radiation}\label{radiation}

Extended exposure of the Observatory to a very high energy radiation
environment can cause components to degrade with total dose.  
For \iMAP the
ambient environment at L$_2$ is not severe, but the passages through the Earth
belts during the early phasing loops can give a significant dose in a short
time.
The predicted total ionizing dose (TID) of radiation exposure at the center 
of a 2.54 mm thick aluminum shell, in \iMAP's trajectory over the 
course of 27 months, is $\approx 13.5$ krad.  For safety, and to cover model 
uncertainties, a factor of two margin was imposed on this 
prediction, so \iMAP was designed to withstand 27 krad TID.  Ray tracing of
the actual \iMAP geometry shows that most electronics boards are exposed 
to about $5-10$ krad.

\iMAP is also designed to a requirement to survive single event upsets at a level of 
$35$ MeV cm$^2$ mg$^{-1}$ linear energy transfer (LET; the energy left 
by an energetic particle), without degraded performance.
  
\subsubsection{Micrometeoroids}

Any object of substantial size in space will be subject to bombardment by
micrometeoroids.  For \iMAP 
micrometeoroids may puncture the MLI webbing between the solar
shields (allowing a limited amount of sunlight to illuminate 
the optics and diffract into the feeds) and impact the optics 
(increasing their emissivity).  Holes in the Sun shield would transmit 
solar energy, while holes in the reflectors would emit as blackbody sources
at 70 K, the approximate temperature of a reflector.  For the isotropic distribution 
of micrometeoroids believed to be representative of L$_2$, a solar leak 
through punctured webbing will not produce a net signal larger than 
0.5 $\mu$K at 93 GHz, even in the absence of the diffraction shield.
For the duration of the mission, the offset from random micrometeoroids
hitting the primaries is expected to be smaller than $0.5~\mu$K at the 95\% confidence 
level if the holes are the size of the micrometeoroids.  If the holes
are five times larger than the micrometeoroids that produced them, which
is not out of the question, the offset will be negligible: $<140~\mu$K
in 95\% of the cases.  Damage to the reflectors is dominated by 
rare encounters with 100 $\mu$m size particles.  See \citet{page02}
for a more detailed discussion of the effects of micrometeoroids on the \iMAP optics.

\subsection{Launch and Trajectory}

A Delta 7425-10 expendable launch vehicle places \iMAP into a
$28.7\dg$ inclination near-circular orbit from the 
NASA Kennedy Space Center Eastern Test Range. The Delta third stage 
fires, de-spins, and then separates from \iMAP, placing \iMAP into a highly eccentric orbit
reaching nearly to the lunar distance. The specific energy (energy
per unit mass) of the launch was fixed at C$_3=-2.6$ km$^2$ s$^{-2}$ to 
place a maximal amount of mass into orbit. The Delta vehicle uses four 
graphite epoxy motors (GEMs), a STAR-48 third stage, and a 3.048 m  
composite fairing.  The absolute value of the launch vehicle velocity error 
is $<11.6$ m \persec ($3\sigma$).  \iMAP is attached to the vehicle with 
a 3712C payload adapter fitting.  During the launch the payload 
experiences accelerations of 11.3 g along the thrust axis and 3.5 g 
laterally. 

Three general options were examined for a trajectory to L$_{2}$: a direct
transfer, a lunar-assisted direct transfer, and a lunar-assisted transfer
with Earth-Moon ``phasing loops'' (i.e., highly eccentric Earth
orbits). The final option was selected based on its fuel efficiency, and 
based on its tolerance for initial problems since there are no critically 
important maneuvers that need to be executed for at least 2 or 3 days after launch. The
lunar gravity-assist requires an orbit apogee of approximately the lunar
orbit distance of $\sim $400,000 km with a line of apsides oriented such
that \iMAP passes just behind the Moon at apogee (a ``trailing swingby''). The
most efficient gravity-assist occurs when the Moon approaches the anti-Sun
direction, near full Moon. Phasing loops  
are used to avoid the narrow launch
window otherwise imposed by the lunar-assist and to allow time to correct
launch vehicle errors and gain spacecraft operational experience before critical 
maneuvers must take place. Two through
five loop scenarios are possible, although only 3 and 5 loop cases are accepted
by \iMAP since they have the least risk. 
Mid-course correction(s) 
apply final corrections following the lunar swingby and before the
Observatory attains a Lissajous orbit about the L$_{2}$ point. Fig. \ref{trajectory}
shows a sample \iMAP trajectory to L$_{2}$. The Goddard Space
Flight Center has done relevant previous trajectories for the L$_{1}$ missions {\it WIND}, 
{\it SOHO}, and {\it ACE}.

The natural growth of trajectory errors in orbit about L$_2$ will result in the need to
execute small station-keeping trim maneuvers approximately every three months.
Momentum unloading is accomplished at the same times to minimize 
thermal disruptions to the Observatory.

\subsection{Ground Operations}

Data from the satellite is transferred through NASA's Deep Space Network (DSN), 
to a combined Science and Mission Operations Center (SMOC), located at the 
Goddard Space Flight Center. Very little in the way of science operations 
activities are required for the \iMAP mission due to the survey nature of the
mission and the desire to minimize all disturbances to the Observatory. 
The SMOC monitors the basic health and safety of the Observatory, sends all
commands, and transfers level-0 data (data that passes parity checks and is 
in time-order) to the Office of the \iMAP Experiment's General Archive (OMEGA).

Two mission design features were implemented to assist the operations team.  First, there
is on-board fault protection.   
Software in the attitude control system can detect a fault, switch in redundant hardware, 
and switch to a simple safe-hold mode to buy time for personnel on the ground to determine 
exactly what happened and to correct the problem before returning \iMAP to Observing mode. 
Software in the power system can detect a low voltage or low battery state-of-charge 
and react by restoring proper settings and shedding loads.
Second, the system supplies the ground with error and status measurements and telemetry
to inform the ground that something has gone wrong or
is out of limits.  Generally, the spacecraft informs the ground that something is threatening the
spacecraft health and safety.

\section{DATA ANALYSIS}

The \iMAP Science Team must:
check the data to assure that all operations appear to be proper; 
assure that the instrument is operating optimally; 
check the data, in great detail, for evidence of systematic errors; 
attempt to correct any such errors; and 
calibrate the data and transform it from time-ordered to a map of 
the sky according to \citet{wright96a}. The idea is to guess a map of the sky, and then 
improve the guess iteratively using the differential data.  These
iterations can include polarization maps and a calibration solution.
The map data must also be carefully checked for systematic errors.  
The heart of the data processing task is to place quantitative 
upper limits on potential systematic measurement errors.

\subsection{Systematic Error Analysis}{\label{syserr}}

The raw differential data may be modeled in the form
\begin{eqnarray*}
d(t)  & = & g(t) \left[\Delta T(t) + s(t) + o(t)\right] + n(t) \\
      & = & \left[g_0(t) + g_1(t)\right] \left[\Delta T(t) + s(t) 
             + o_0(t) + o_1(t)\right] + n(t) \\
      & = & g_0(t) \Delta T(t) + g_0(t) s(t) + g_0(t) o_0(t) \\
      &   & +\, g_0(t) o_1(t) + g_1(t) o_0(t) + n(t) + \cdots
\end{eqnarray*}
where $d(t)$ is the raw time-stream in uncalibrated digital units, 
$g(t)$ is the true radiometer gain, in du mK$^{-1}$, $\Delta T(t)$ is the true 
differential sky temperature, in mK,  $s(t)$ is the signal due to uncorrected 
differential signals, in mK, that act on the instrument sidelobes, $o(t)$ is 
the instrument offset, in mK, which is the signal that would be measured with 
a nulled input signal, and $n(t)$ is the instrument noise produced either by 
the amplifiers, or by miscellaneous pick-up after signal amplification.  It is
useful to expand the gain and offset in a form of perturbation series, where
$g_0(t)$ and $o_0(t)$ are the dominant, slowly varying terms and $g_1(t)$ and 
$o_1(t)$ are the smaller, more rapidly fluctuating terms.  Terms of order 
$g_1 o_1$ are dropped in the last line of the above expression.

As part of the data processing, known modulation due to the CMB
dipole is used to fit for an instrument gain and baseline (see \S\ref{cal}).  Noise levels 
are such that reasonable sensitivity (to better than 1\%) to each term
is achieved with about an hour of data.  
Since this matches the spacecraft precession 
period, it marks a characteristic time scale for the above perturbation 
expansion.  Gain and offset changes can be tracked on time scales 
greater than one hour directly from the sky data, while other 
means must be employed to track changes on shorter time scales, 
particularly at the spin period 
since these changes can most closely mimic a true sky signal.

In processing the data, smooth fits to the gain and baseline solutions are 
generated, and are denoted 
the recovered gain and baseline, $g'(t)$ and $b'(t)$, respectively.  The 
recovered differential temperature is then
\begin{eqnarray*}
\Delta T'(t)& = & \left[d(t) - b'(t)\right]/g'(t) \\
            & = & \left[g_0 \Delta T + g_0 o_0 + g_0 o_1 + g_0 s\right. \\
            &   & \left.+\, g_1 o_0 + n - b'\right]/g'  + \cdots \\
            & = & (g_0/g') \Delta T + \left[(g_0/g') o_0 - b'/g'\right] \\
            &   &  +\, (g_0/g') o_1 + (g_0/g') s + (g_1/g') o_0 + n/g' + \cdots
\end{eqnarray*}
where the explicit time dependence in each term is dropped.  Note that the
true baseline and the true offset are related by the gain, $b(t) = g(t) o(t)$.

Systematic errors in the final sky maps can originate from a variety of 
sources that can be classified according to which term they contribute to in 
the above expansion.  These include:
\begin{itemize}

\item Calibration errors.  These calibration errors, from any source, 
contribute to making $(g_0/g')$ differ from unity and to making 
$(g_0/g') o_0 - b'/g'$ differ from zero. (see \S\ref{cal})

\item External emission sources.  These include errors due to spin-synchronous 
modulations of the emission from the Sun, Earth, Moon, and Galaxy acting on the 
instrument sidelobes, or due
to the local Doppler effect producing an induced signal.  These contribute to 
$(g_0/g') s$ (see \S\ref{extem}).

\item Internal emission sources.  These errors are due to spin-synchronous 
temperature variations acting on components with fixed emissivities.  These 
effects contribute to $(g_0/g') o_1$ by varying the instrument offset 
(see \S\ref{stability}).

\item Multiplicative electronics sources.  These errors are due to spin synchronous 
gain variations acting on a fixed radiometric offset.  These effects contribute 
to $(g_1/g') o_0$ (see \S\ref{pduaeu}).

\item Additive electronics sources.  These errors are due to miscellaneous 
spin-synchronous electronics errors, such as instrument channel-channel 
cross-talk.  These contribute to $n(t)$, which needn't be random, or white 
(see \S\ref{radiometer}).

\item Striping.  These effects can be introduced by correlations in the 
instrument noise, due either to 1/f effects or to post-detection filtering,  
or due to the effects of scan smearing and intrinsic beam ellipticity.

\item Map-making errors. These errors are due to poor convergence or striping
introduced by the map-making algorithm, possibly in concert with calibration
errors.  Also errors due to pointing uncertainty.

\item Beam mapping errors. These are errors in the determination of the 
main beam window function that directly contribute to errors in the recovered 
power spectrum.

\end{itemize}

\subsection{Calibration Analysis \label{cal}}

For a sufficiently short period of time the instrument gain and baseline can be
approximated as constant,
\begin{equation}
d(t) \approx g_n \Delta T(t) + b_n
\end{equation}
where $g_n$ and $b_n$ are the gain and baseline during the $n^{\rm th}$ calibration
period.  Since the sky signal $\Delta T$ is dominated by the (known) CMB dipole 
$\Delta T_d$ (including the time-dependent modulation from \iMAP's velocity relative 
to the Sun), the raw data for a gain and a baseline can be fit by minimizing 
\begin{equation}
\chi^2 = \sum_i \frac{\left(d(t_i) - (g_n \Delta T_d(t_i) + b_n)\right)^2}
                     {\sigma_i^2}
\end{equation}
To minimize the covariance between the recovered gain and the baseline
(offset) it is necessary to have a scan strategy such that the time average 
of $\Delta T_d$ is nearly zero.  The \iMAP combined spin and precession 
is designed to produce a scan strategy that satisfies this requirement.

Any difference in the recovered gain or
baseline compared to the slowly varying components $g_0(t)$ and $g_0(t)o_0(t)$,
respectively, is defined as a calibration error.  
These differences are most easily tracked by end-to-end 
simulations in which raw differential data is simulated with a known input 
signal and calibration.  These data can be run through the pipeline that
calibrates the data and solves for the sky map and the recovered 
calibration can
be compared to the known inputs.  The corresponding effects on the sky maps can
be inferred by comparing the recovered maps to the known inputs.

There are a number of effects that can cause calibration errors:
\begin{itemize}
\item Instrument noise -- the recovered gain and baseline will have random
errors due to instrument noise.  This is typically less than 1\% per hour 
per channel of data.  This can be further reduced with filtering matched to the 
specific properties of the individual radiometers.
\item Anisotropy -- higher-order CMB or galactic anisotropy, $\Delta T_a = \Delta T - 
\Delta T_d$ can significantly project onto the dipole over the course of any 
given calibration period, so the calibration fit is iteratively improved by 
subtracting an estimate of the anisotropy from the raw data prior to fitting.  
In particular, one can minimize the modified $\chi^2$
\begin{equation}
\chi^2 = \sum_i \frac{\left(d'(t_i) - (g_n \Delta T_d(t_i) + b_n)\right)^2}
                     {\sigma_i^2}
\end{equation}
where
\begin{equation}
d'(t_i) = d(t_i) - {\tilde g}_n \Delta{\tilde T}_a(t_i)
\end{equation}
and where ${\tilde g}_n$ is an estimate of the gain from a previous calibration 
iteration, and $\Delta{\tilde T}_a(t_i)$ is an estimate of the anisotropy from 
a previous sky map iteration.
\item Dipole uncertainty -- The absolute calibration is determined using the modulation of 
the CMB dipole due to the motion of \iMAP with respect to the Sun.  The 0.7\% uncertainty in
the {\it COBE} dipole is removed by the anisotropy correction described above.

\end{itemize}

Additional improvements in calibration accuracy may be possible beyond the basic process 
outlined above.  For example, instrument 
house-keeping telemetry may be used to provide independent tracking of the relative gain 
of the instrument.  Iterative refinements in the basic algorithm are also possible.  
Flight data will be used to explore the possibilities.

\subsection{External emission sources \label{extem}}

Emission from the Sun, Earth, Moon, and Galaxy can contaminate the raw data by
entering the instrument via the sidelobes of the beams.  In the case of emission from
the Sun, Earth, and Moon this can only occur after the signal diffracts around
the solar array shield.  In the case of the Galaxy only that emission
which enters via the sidelobes at $>5^\circ$ from the boresight is considered a 
systematic error.  Galactic emission that enters in or near the main beam is
considered a foreground signal and is treated using multi-frequency sky map analysis
(see \S\ref{multifreq}).

Another source of external emission is the dipole signal induced by \iMAP's motion
with respect to the Sun. (The portion due to to the Sun's motion with respect to
the CMB rest frame is treated separately).  Since this signal is used as an
absolute calibration source, we treat this effect as a calibration error.

The data analysis pipeline reads and calibrates raw
differential data, corrects the data for known sky signals and systematic
effects, and bins the results by the pixel number of the source direction in
spacecraft coordinates.  This produces a differential beam map over the 
portion
of the sphere in which the given source is visible.  In the case of the Sun,
Earth, and Moon, the portion of the sphere covered from L$_2$ will be limited 
to a cone about the +z direction.  (Far-sidelobe measurements using the Moon may
be possible from pre-L$_2$ the early flight operations.)
In the case of the Galaxy, the beam map will cover the
full sky, but it will only serve as a cross check of the full sidelobe maps
measured on the ground.

\subsection{Data products}

All \iMAP scientific data will be validated and then released via NASA's newly created 
cosmic microwave background data center: the Legacy Archive for Microwave Background 
Data Analysis (LAMBDA).  The initial data release is expected approximately 18 months 
after launch.  Subsequent data deliveries will come in stages, with logarithmic time 
interviews, in a manner similar to the {\it COBE} mission's delivery of anisotropy data 
(i.e., 1-year, 2-year, 4-year).  \iMAP is currently approved to operate for 4-years.

\acknowledgements

The \iMAP mission is made possible by the support of the Office of Space 
Sciences at NASA Headquarters and by the hard and capable work of scores of 
scientists, engineers, technicians, machinists, data analysts, budget analysts, 
managers, administrative staff, and reviewers.

\end{document}